\documentclass[preprint,12pt,authoryear]{elsarticle}
\usepackage[margin=2.5cm]{geometry}
\usepackage{enumitem}
\newlist{inlinelist}{enumerate*}{1}
\setlist*[inlinelist,1]{%
  label=(\roman*),
}
\usepackage{soul}
\usepackage{hhline}
\usepackage{subfig}
\usepackage{multirow}
\usepackage{xcolor}
\usepackage{graphicx}
\usepackage{upgreek}
\usepackage{amssymb}
\usepackage{amsfonts,amsthm,bm,amsmath} 
\usepackage{appendix}
\usepackage{natbib}
\setcitestyle{sort&compress}
\usepackage{times}
\usepackage{caption}
\usepackage{subfig}
\newcommand{\psubref}[1]{\protect\subref{#1}}

\usepackage{lineno}

\usepackage[colorlinks]{hyperref} 
\hypersetup{ 
    colorlinks=true,       
    linkcolor=red,          
    citecolor=blue,        
    filecolor=magenta,      
    urlcolor=cyan           
}

\newcommand{\fref}[1]{Fig.~\ref{#1}}

\newcommand{\eref}[1]{Eq.~(\ref{#1})}
\newcommand{\erefs}[2]{Eqs.~(\ref{#1})--(\ref{#2})}
\newcommand{\sref}[1]{Section~\ref{#1}}
\newcommand{\tref}[1]{Table~\ref{#1}}

\journal{Modelling and Simulation in Materials Science and Engineering}
\begin{document}

\begin{frontmatter}

\title{Polycrystal plasticity with grain boundary evolution: A
    numerically efficient dislocation-based diffuse-interface model}

\author[]{Junyan He}
\author[]{Nikhil Chandra Admal\corref{mycorrespondingauthor}}
\cortext[mycorrespondingauthor]{Corresponding author}
\ead{admal@illinois.edu}

\address{Department of Mechanical Science and Engineering}
\address{University of Illinois at Urbana-Champaign, Champaign, IL, USA}

\begin{abstract}
Grain structure plays a key role in the mechanical properties of alloy materials. Engineering the grain structure requires a comprehensive understanding of the evolution of grain boundaries (GBs) when a material is subjected to various manufacturing processes. To this end, we present a
computationally efficient framework to describe
the co-evolution of bulk plasticity and GBs. We represent GBs as diffused
geometrically necessary dislocations, whose evolution describes GB plasticity.
Under this representation, the evolution of GBs and bulk plasticity is
described in unison using the evolution equation for the plastic deformation
gradient, an equation central to classical crystal plasticity theories. To reduce the number of degrees of freedom, we present a procedure which combines the governing equations for each slip rates into a set of governing equations for the plastic deformation gradient. Finally, we outline a method to introduce a synthetic potential to drive migration of a flat GB.

Three numerical examples are presented to demonstrate the model. First, a scaling test is used to demonstrate the  computational efficiency of our framework. Second, we study the evolution of a tricrystal, formed by embedding a circular grain into a bicrystal, and demonstrate qualitative agreement between the predictions of our model and those of molecular dynamics simulations by \cite{trautt2014capillary}. Finally, we demonstrate the effect of applied loading in texture evolution by simulating the evolution of a synthetic polycrystal under applied displacements.

\end{abstract}

\begin{keyword}
Polycrystal plasticity \sep Diffuse-interface model \sep Grain boundary migration
\end{keyword}

\end{frontmatter}

\section{Introduction}
\label{sec:intro}
The grain microstructure of polycrystalline materials plays a significant role
in their mechanical performance \citep{may2007mechanical,korner2014tailoring}.
However, it is not a stationary quantity --- both hot and cold working processes
change the underlying grain structure, thus altering the mechanical properties
\citep{giessen1967crystal,senkov2018effect}. Advancements in modern manufacturing
techniques have allowed accurate control of the manufacturing parameters, and
the idea of grain boundary (GB) engineering \citep{watanabe2011grain} has gained
momentum ever since. The idea underpinning GB engineering is to optimize the GB
structure to achieve desired mechanical properties, through various
manufacturing processes and treatments \citep{watanabe2011grain}. However,
effective GB engineering relies on a comprehensive understanding of the
process-structure relationship for various manufacturing processes. That is, one
has to be able to predict the GB evolution given the thermomechanical loads in a manufacturing process. 

It is well known from experiments \citep{li1953stress,bainbridge1954recent} that as a GB migrates, the underlying material undergoes plastic deformation. The coupling between GB motion and the accompanying plasticity is commonly referred to as \emph{grain boundary coupling}. In a coupled GB motion, the normal motion of a GB is accompanied by a translation tangential to the GB plane, and the
extent of coupling is measured using the \emph{coupling factor} $\beta$ \citep{cahn2006coupling}, defined as
\begin{equation*}
    \beta = \frac{v_t}{v_n},
\end{equation*}
where $v_n$ and $v_t$ are the normal and tangential velocities of the GB.

An important consequence of GB coupling in the presence of curvature is the possibility of grain rotation \citep{mishin2010atomistic}, which implies GB migration sometimes results in misorientation changes, which in turn alter the properties of GBs. Recent atomic scale studies \citep{thomas2017reconciling, han2018grain} have shown that the coupling factor of a GB is not a fixed quantity, but depends on multiple factors such as temperature and the nature of loading, in addition to the misorientation and inclination of the GB. 
Recognizing the chain of causality in coupled GB motion --- the incompatibility of the plastic deformation accompanying GB motion results in internal stresses, which in turn contributes toward the driving forces on GBs --- conveys the complexity of GB motion. Therefore, recent works have adopted a coupled approach, wherein GBs and the induced plastic shear deformation co-evolve. We will broadly refer to models wherein an intrinsic, but possibly non-constant, plastic shear deformation is associated with GB motion as \emph{shear-coupled GB models}. We emphasize that there are models \citep{abrivard2012phase,zhao2016integrated,jafari2019modeling,mikula2019phase} that describe the co-evolution of bulk plasticity and GBs, but do not account for the intrinsic plastic shear deformation associated with GB motion. In what follows, we limit our literature survey to shear-coupled GB models.

Existing shear-coupled GB models broadly differ in their modeling of the intrinsic GB plasticity and deformation. In \citet{basak2014two,basak2015simultaneous}, the authors assume the grains undergo rigid rotation as a response to shear coupling and sliding, and the necessary plastic shape change is enabled by surface diffusion. Moreover, GBs are assumed to be sharp interfaces with anisotropic GB energy and an associated coupling factor. On the other hand, \citet{zhu2014continuum,zhang2018motion,zhang2020new} model sharp-interface GBs using lattice dislocations which not only give rise to GB energy but also to an intrinsic GB plasticity. Grain boundaries evolve as a response to the short and long range stress fields computed using the theory of dislocations. 
It is important to note that deformation is not modeled explicitly in the above models.
On the other hand, the model by \citet{admal2018unified} incorporates shear-coupled GB motion using a unified framework that describes bulk and GB plasticity simultaneously within the theory of crystal plasticity. The central idea here is to construct GBs using, in the language of \cite{MURA1989149},  \emph{impotent} geometrically necessary dislocations (GNDs) that generate no elastic stresses but result in the lattice misorientation corresponding to a GB. Using a constitutive law enriched with a GB energy that is a function of the grain boundary GNDs, \citet{admal2018unified} have demonstrated that the resulting model can describe phenomena such as shear-coupled GB motion, GB sliding, grain rotation and recovery. However, the model is computationally expensive as the number of coupled differential equations scales with the number of slip systems, limiting the study to simple bi- and tri-crystals in two dimensions. In addition, the model does not include the effect of statistically stored dislocations (SSDs), a useful characteristic as demonstrated by \citet{ask2019cosserat,mikula2019phase}.
The above-mentioned challenges motivate the main objectives of this paper, which are to reformulate the model of \citet{admal2018unified} to make it computationally tractable, and extend our study of GB plasticity from bicrystals to polycrystals. In addition, we equip the model with a synthetic driving force for GB migration which emulates a chemical potential in discrete systems.

In this paper, we also discuss the limitations of the dislocations-based framework of \citet{admal2018unified}. In this regard, we take note of the recent development of a new class of shear-coupled GB models \citep{thomas2017reconciling,thomas2019disconnection,wei2019continuum,gokuli2021multiphase,runnels2020phase} developed in the backdrop of overwhelming evidence from atomic scale simulations or experiments \citep{zhu2019situ,merkle2002thermally,doi:10.1080/14786435.2012.760760,PhysRevLett.110.265507,mompiou2015situ} that \emph{disconnections} are the primary carriers of GB plasticity. Disconnection-based continuum models, in which GB motion is dictated by the evolution of continuum disconnections, offer unique advantages compared to our dislocation-based framework. However, since a disconnection-based framework describes only a discrete collection of GBs with a well-defined coincident site lattice, we discuss how the two approaches can potentially complement each other to model GBs with arbitrary misorientations.

This paper is organized as follows. In \sref{sec:formulation}, we present a diffuse-interface framework to model GB plasticity that is computationally tractable and equipped with a synthetic potential that emulates stored energy due to SSDs. In \sref{sec:numerics}, we drive GB motion with a synthetic potential, demonstrate the computational efficiency with a scaling test, and compare our continuum simulations to MD simulations of previous works of \citet{trautt2014capillary}. In addition, we simulate GB plasticity in a polycrystal, and demonstrate the effect of external loads on GB evolution and grain rotation. In \sref{sec:conclusion}, we conclude by outlining the limitations and discuss on the need to combine a disconnection- and dislocation-based models to better describe migration of arbitrary GBs. Standard continuum mechanics notation is adopted unless otherwise noted.

\section{A diffuse-interface framework for modeling grain boundary plasticity in polycrystals}
\label{sec:formulation}
In this section, we develop a diffuse-interface framework to model grain boundary plasticity that is computationally more efficient than the one developed by \citet{admal2018unified}. We begin with a description of the framework of \citet{admal2018unified} that is based on crystal plasticity for polycrystals equipped with grain boundary energy. Next, we highlight the features that make it computationally expensive. We then proceed to derive an alternative formulation that results in fewer degrees of freedom (DOF) per node and an increase in computational efficiency. Finally, we propose a synthetic driving force for GB motion within the model, which is analogous to the chemical potential approach developed by \cite{janssens2006computing} for driving GBs in molecular dynamics (MD) simulations.

\subsection{Kinematics}
\label{sec:kinematics}
A body is represented by an open subset $B$ of a Euclidean
space $\mathbb R^3$. A time-dependent motion of $B$, described relative to a reference configuration $B_0 \subset \mathbb R^3$, is given by a smooth one-to-one map $\bm{x}(\bm{X},t)$, which maps a material point $\bm X$ to its spacial position $\bm x$ at time $t$. Let $\bm{F}(\bm{X},t) := \nabla\bm{x}(\bm{X},t)$ denote the gradient of the deformation map at time $t$, where $\nabla$ denotes the gradient with respect to the material coordinate $\bm X$.

In finite deformation plasticity, the
deformation gradient admits a multiplicative decomposition
\citep{hill1966generalized,lee1969elastic,reina2014kinematic}
\begin{equation}
    \bm{F} = \bm{F}^{\rm L}\bm{F}^{\rm P},
    \label{E:1}
\end{equation}
where $\bm{F}^{\rm L}$ and $\bm{F}^{\rm P}$ denote the lattice\footnote{Also
    known as the elastic distortion, $\bm{F}^{\rm E}$. The notation
    $\bm{F}^{\rm L}$ is adopted from \citet{clayton2010nonlinear}.} 
and plastic
distortions, respectively. $\bm{F}^{\rm P}$ maps an infinitesimal line element
$d\bm{X}$ to $\bm{F}^{\rm P}d\bm{X}$. If we regard the collection of line
elements $\bm{F}^{\rm P}d\bm{X}$ as the lattice configuration, then $\bm{F}^{\rm
P}$ can be understood as a linear transformation from the reference configuration to the
lattice configuration, whereas $\bm{F}^{\rm L}$ is a linear mapping from the lattice
configuration to the deformed configuration \citep{admal2018unified}.\footnote{Strictly speaking, $\bm F$, $\bm F^{\rm L}$ and $\bm F^{\rm P}$ are linear transformations between the tangent spaces of the respective configurations.}
In this paper, the body $B$ represents a polycrystalline material, and we limit the mechanism of plastic deformation to dislocation slip on the local crystal's slip systems. Therefore, assuming the crystal has $N_s$ slip systems, the evolution of $\bm{F}^{\rm P}$ is given by 
\begin{equation}
    \dot{\bm{F}^{\rm P}} = \bm{L}^{\rm P}\bm{F}^{\rm P} = \left (\sum_{\alpha=1}^{N_s}\nu^\alpha\bm{S}^\alpha \right) \bm{F}^{\rm P},
    \label{eqn:flow}
\end{equation}
where $\bm{L}^{\rm P}$ denotes the plastic velocity gradient given in terms of the Schmid tensors $\bm S^\alpha$ and the slip rates $\nu^\alpha$ ($\alpha=1,\dots,N_s$). The Schmid tensor $\bm S^\alpha:=\bm{s}^\alpha\otimes\bm{m}^\alpha$ is defined in terms of the $\alpha$-th slip direction and slip plane normal, $\bm s^\alpha$ and $\bm m^\alpha$, respectively.

\subsection{Grain boundaries expressed as geometrically necessary dislocations}
\label{sec:gnd}
In order to model evolving grain boundaries within a crystal plasticity framework, \citet{admal2018unified} proposed a construction of diffuse-interface grain boundaries using GNDs, followed by equipping their model with a GB energy density in terms of the GND density. We will now describe their construction of GNDs that results in a stress-free polycrystal.

Traditional crystal plasticity theories describe the deformation of a polycrystal using a
strain-free \emph{polycrystal} as a reference configuration, along with an initial decomposition ($\bm F=\bm F^{\rm E} \bm F^{\rm P}$) of the deformation gradient given by\footnote{
Notice that the notation $\bm F^{\rm E}$ is used in \eref{eqn:split_old} instead of $\bm F^{\rm L}$. This is because $\bm F^{\rm E}$ describes the elastic distortion of the lattice relative to the reference polycrystal, and does not include the orientation of the grain.}
\begin{equation}
    \bm{F}(\bm{X},0) = \bm{F}^{\rm E}(\bm{X},0) = \bm{F}^{\rm P}(\bm{X},0) =
    \bm{I}.
    \label{eqn:split_old}
\end{equation}
Since the reference configuration is a polycrystal, the Schmid tensors, defined in \sref{sec:kinematics}, are piecewise-constant and depend on the orientation of each grain. While the multiplicative decomposition of $\bm F=\bm I$ in \eref{eqn:split_old} is a reasonable choice for modeling polycrystal plasticity with fixed grain boundaries, it does not lend itself to a construction of GB energy and modeling evolving GBs.

An alternate approach is to choose a strain-free \emph{single} crystal as the reference configuration, along with an initial decomposition 
\begin{equation}
    \bm{F}^{\rm L}(\bm{X},0) = \bm{R}^0(\bm{X}); \; \bm{F}^{\rm P}(\bm{X},0)
    =\bm{R}^0(\bm{X})^T,
    \label{eqn:split_new}
\end{equation}
where $\bm{R}^0(\bm{X}) \in SO(3)$ is a piecewise-constant rotation field
indicating the crystal orientation of each grain. By construction, $\bm F(\bm X,0)\equiv \bm I$, and the initial Green--Lagrangian lattice strain
$\bm E^{\rm L}(\bm X,0) =  ((\bm{F}^{\rm L})^T\bm{F}^{\rm L} - \bm{I})/2 = \bm
0$, which implies the resulting polycrystal is strain-free. At the atomistic scale, it is well known that the dislocation cores of the low-angle GBs result in lattice distortion and stress fields, hence the GB is not strain free. However, the stress fields are short-ranged, and the associated elastic strain energy is already accounted for in the form of GB energy. Hence, in the continuum scale, we model GBs with \emph{impotent} GNDs that do not generate any stresses, which results in a Read--Shockley type GB energy.
Moreover, since the reference configuration is a single crystal, the Schmid tensors are now constant, and correspond to the slip systems of the single crystal.
For numerical convenience, the piecewise constant field $\bm R^0(\bm X)$ is
replaced with a smoothed version $\tilde{\bm{R}}^0(\bm{X}) \in SO(3)$, leading to a diffuse-interface model. 

The main advantage of the decomposition given in \eref{eqn:split_new} is that GBs are characterized by the GND density, defined as
\begin{equation}
    \bm{G} = \bm{F}^{\rm P} {\rm Curl } (\bm{F}^{\rm P}),
    \label{eqn:gnd}
\end{equation}
where Curl denotes the curl\footnote{In indicial notation, $({\rm{Curl}}
\bm{A})_{ij} = \epsilon_{ipq}A_{jq,p}$.} of a tensor field in material
coordinates. The GND density tensor can then be used to construct the grain
boundary contribution of the free energy functional, as shown in the next
section. 

\subsection{A constitutive law to describe bulk and grain boundary energy}
\label{sec:energy}
In this section, we begin by describing the constitutive law proposed by \citet{admal2018unified} that results in a simultaneous evolution of grain boundary mediated phenomena and bulk plasticity. The construction of the constitutive law for the GB energy is based on the assumption that the GND density that constitutes the GB quantifies its energy. Therefore, the total energy density $\psi$ is given by a sum of a bulk elastic energy density $\psi^{\rm e}(\bm E^{\rm L})$, and a  GB energy density $\psi^{\rm gb}$, i.e.
$\psi = \psi^{\rm e} + \psi^{\rm gb}$. For simplicity, we limit ourselves to isothermal condition.

The elastic energy density is assumed to be of the classical form that depends only on the lattice strain $\bm{E}^{\rm L}$. On the other hand, the construction of the GB energy density is inspired by the KWC model \citep{kobayashi1998vector}, a diffuse-interface model for grain microstructure evolution with two order parameters $\phi$ and $\theta$ that describe the degree of crystallinity and crystal orientation, respectively. The GB energy density of the KWC model is given by
\begin{equation}
    \psi^{\rm KWC}(\phi,\nabla\phi,\nabla\theta) = \frac{\alpha^2}{2}|\nabla\phi|^2 + e_0 f(\phi) +
    sg(\phi)|\nabla \theta| + \frac{\epsilon^2}{2}|\nabla \theta|^2,
    \label{eqn:gbEnergy_kwc}
\end{equation}
where 
\begin{equation}
    f(\phi) = (\phi-1)^2, \quad
    g(\phi) = -2({\rm ln}(1-\phi)-\phi).
    \label{eqn:functions}
\end{equation}
The parameters $\alpha$, $e_0$, $s$, and $\epsilon$ are material parameters that
determine the GB energy and width. The construction of
    $\psi^{\rm KWC}$ by \citet{kobayashi2000continuum} was centered around the
    observation that the term $sg(\phi)|\nabla\theta|$ acts to localize the
    grain boundary, while the term $\frac{\epsilon^2}{2}|\nabla\theta|^2$ tends
to diffuse it. The combined effect of the two gives rise to GBs of finite widths. Moreover, the inclusion of the $|\nabla\theta|^2$ term results in motion by curvature. Noting that GNDs measure the gradient of lattice rotation, and the motion of GBs is given by the evolution of GNDs, \citet{admal2018unified} have shown that by  replacing $\nabla \theta$ with $\bm G$, and defining $\psi^{\rm gb}$ as
\begin{equation}
    \psi^{\rm gb} = \frac{\alpha^2}{2}|\nabla\phi|^2 + e_0 f(\phi) +
    sg(\phi)|\bm{G}| + \frac{\epsilon^2}{2}|\bm{G}|^2,
    \label{eqn:gbEnergy}
\end{equation}
where $|\bm{G}|$ denotes the standard Euclidean norm of the GND density tensor, the total energy
\begin{equation}
    W[\phi,\bm u,\bm F^{\rm p}] = \int_{B_0} \psi^{\rm e}(\bm E^{\rm L}) + \psi^{\rm gb}(\phi,\nabla \phi,\bm G) \, dV,
    \label{eqn:W}
\end{equation}
is a functional of $\phi$, the displacement field $\bm u:=\bm x(\bm X,t)-\bm X$, and $\bm F^{\rm P}$. The resulting model
not only inherits all the features of the KWC model, such as GB motion by curvature and grain rotation, but it also describes additional phenomena, such as coupled GB motion, sliding and recovery. This is because the GNDs constituting the GBs respond to stress,\footnote{Currently, we limit this response to dislocation glide.} and the shear accompanying their glide results in shear-induced grain boundary motion. \emph{It is important to recognize that, depending on the availability of slip systems and the loading condition, the slip rates in a GB may also result in a combination of sliding and pure coupling.}

We now extend the constitutive law given in \eref{eqn:gbEnergy_kwc} to include  energy due to SSDs, which play an important role in the evolution of grain microstructure of cold-formed polycrystals. Our construction of the energy due to SSDs is inspired by the synthetic potential energy for atomistic systems proposed by \citet{janssens2006computing}, where the motivation was to induce GB motion in bicrystals in the absence of curvature.\footnote{Although a flat grain boundary can be driven by shear stress, this approach is not preferred in MD simulations because the limited times scales accessible to MD necessitate the use of extremely high stress and strain rates that do not represent commonly observed deformation rates.}
Selectively adding potential energy to atoms on one side of the
GB, results in an artificial force on the atoms with higher potential which 
reorients them with atoms on the other side of the GB. We will now present a continuum analog of the synthetic potential, which represents energy due to SSDs. We limit the description to symmetric tilt grain boundaries in two dimensions.

A synthetic potential is incorporated into $\psi^{\rm{gb}}$ by modifying the functional form of $f(\phi)$ in \eref{eqn:functions} to
\begin{equation}
    f(\phi) = (\phi-1)^2 + \Delta E^*  [ 1 - {\rm{tanh}}( \rho \theta^{\rm{L}}) ] \phi,
    \label{eqn:modification}
\end{equation}
where $\rho$ denotes a
user-defined scaling parameter, and $\theta^{\rm{L}}$ denotes the current crystal orientation field, which can be obtained from the lattice rotation $\bm R^{\rm L}$ computed using the right polar decomposition of $\bm F^{\rm L}$. The term $[1 - {\rm{tanh}}(\rho
\theta^{\rm{L}})]$ can be interpreted as a grain selector function, which serves to allocate an additional energy of $e_0 \Delta E^*$ to the selected grain (in this case, the grain with a negative $\theta^{\rm{L}}$). In \sref{sec:scale}, we demonstrate a numerical implementation of a GB driven by synthetic potential.

\subsection{Governing equations}
\label{sec:equations}
One of the main goals of this paper is to formulate a computationally tractable model that can model GB and bulk plasticity. In this section, we derive evolution equations for the unknown fields --- displacement $\bm u:= \bm x -\bm X$, plastic distortion $\bm F^{\rm P}$, and the order parameter $\phi$ --- that are computationally more tractable than those derived by \citet{admal2018unified}. In \sref{sec:numerics}, we demonstrate the computational efficiency of our formulation. 

Our derivation is based on the virtual work formulation of \cite{gurtin2008finite}. The principle of virtual work states that the power expended on an arbitrary part of a body by external forces is equal to the internal power expended within that part by internal forces. We assume that the internal power, denoted by $I(P)$, is expended on a part $P \subset B_0$ by a stress tensor $\bm{P}$ power conjugate to $\dot{\bm{F}}$, a stress vector $\bm{p}$ conjugate to $\nabla\dot{\phi}$, microstress vectors $\bm{\xi}^\alpha$ conjugate to $\nabla\nu^\alpha$, and internal forces\footnote{Internal forces contribute neither to the work nor to the second law.} $\pi$ and $\Pi^\alpha$ conjugate to $\dot{\phi}$ and slip rate $\nu^\alpha$, respectively. Therefore, $I(P)$ is given by
\begin{equation}
    I(P) = \int_{P} \left[ ( \bm{P}:\dot{\bm{F}} + \bm{p} \cdot \nabla\dot{\phi} + \pi \dot{\phi} ) + \sum_{\alpha=1}^{N_s}(\Pi^\alpha\nu^\alpha + \bm{\xi}^\alpha \cdot \nabla\nu^\alpha) \right] dV.
    \label{E:int_pow}
\end{equation}
Using the divergence theorem, \eref{E:int_pow} transforms to 
\begin{align}
    I(P) &= \int_{\partial P} \left[ \bm{P}\bm{N}\cdot\dot{\bm{y}} + \bm{p}\cdot\bm{N}\dot{\phi} + \sum_{\alpha=1}^{N_s}(\bm{\xi}^\alpha \cdot \bm{N})\nu^\alpha \right] dA \notag \\
     &+ \int_{P} \left[ -{\rm{Div}}\, \bm{P}\cdot\dot{\bm{y}} + (\pi - {\rm{Div}}\, \bm{p})\dot{\phi} + \sum_{\alpha=1}^{N_s}(\Pi^\alpha -{\rm{Div}}\,\bm{\xi}^\alpha )\nu^\alpha \right] dV,
  \label{E:int_pow_expanded}
\end{align}
where $\bm N$ denotes the outward unit normal to $\partial P$.
The boundary integral in \eref{E:int_pow_expanded} suggests the following form for the power expended on $P$ by external forces $\bm t$, $s$ and $\bm \Xi$ that are power conjugate to $\dot{\bm y}$, $\dot\phi$ and $v$, respectively:
\begin{equation}
    E(P) = \int_{\partial P} \left[ \bm{t}(\bm{N})\cdot\dot{\bm{y}} + s(\bm{N})\dot{\phi} + \sum_{\alpha=1}^{N_s}\bm{\Xi}^\alpha(\bm{N})\nu^\alpha \right] dA.
    \label{E:ext_pow}
\end{equation}
Enforcing $I(P)=E(P)$ for all $P\subset B_0$, we arrive at the following balance equations
\begin{itemize}
\item  The standard force balance equation for $\bm u$
\begin{equation}
  \begin{array}{l}
    {\rm{Div}} \, \bm{P} = \bm{0} \quad {\rm in} \; B_0, \\ 
    \bm{t} = \bm{P}\bm{N} \;\;\;\;\;\; {\rm on} \; \partial B_0.
  \end{array}
    \label{eqn:u}
\end{equation}
\item A microscopic force balance for $\phi$
\begin{equation}
  \begin{array}{l}
    {\rm{Div}} \, \bm{p} - \pi = 0 \;\;\;\; {\rm in} \; B_0, \\ 
    s = \bm{p}\cdot\bm{N} \;\;\;\;\;\;\;\;\;\; {\rm on} \; \partial B_0.
  \end{array}
    \label{eqn:phi}
\end{equation}
\item A microscopic force balance on each slip system for $\nu^\alpha$
\begin{equation}
  \begin{array}{l}
    {\rm{Div}} \, \bm{\xi}^\alpha - \Pi^\alpha = 0 \;\;\; {\rm in} \; B_0, \\ 
    \Xi^\alpha = \bm{\xi}^\alpha\cdot\bm{N} \;\;\;\;\;\;\;\;\; {\rm on} \; \partial B_0.
  \end{array}
    \label{eqn:v}
\end{equation}
\end{itemize}
Next, we obtain the following constitutive relations between stresses and kinematic fields by examining the first and second laws of thermodynamics using the Coleman--Noll procedure (see \ref{sec:appendix})
\begin{align}
    \bm{P} &= \bm{F}^{\rm L}\frac{\partial\psi^{\rm e}}{\partial\bm{E}^{\rm L}}(\bm{F}^{\rm P})^{-T}, \label{eqn:pk}\\
    \bm{p} &= \frac{\partial\psi}{\partial\nabla\phi},
    \label{eqn:vector_p}\\
    \bm{\xi}^\alpha &= ((\bm{F}^{\rm P})^T\bm{m}^\alpha)\times \left( (\bm{F}^{\rm
    P})^T \frac{\partial\psi^{\rm{gb}}}{\partial\bm{G}}\bm{S}^\alpha \right) + B^\alpha\nabla\nu^\alpha \notag \\
    &=: \bm\xi^\alpha_{\rm e} + \bm\xi^\alpha_{\rm d},
    \label{eqn:xi}
\end{align}
where we split the microstress $\bm \xi^\alpha$ into an energetic part $\bm \xi^\alpha_{\rm e}$ and a dissipative part $\bm \xi^\alpha_{\rm d}$, and  $B^\alpha>0$ is the inverse mobility for $\nabla v^\alpha$. 
The internal forces $\pi$ and $\Pi^\alpha$ are given by
\begin{align}
    \pi &= \frac{\partial\psi^{\rm gb}}{\partial\phi} + b^\phi \dot{\phi},
    \label{eqn:scalar_pi}\\
    \Pi^\alpha &= \frac{\partial\psi^{\rm{gb}}}{\partial\bm{G}}:(\bm{S}^\alpha\bm{G} +
    \bm{G}(\bm{S}^{\alpha})^T) - \tau^\alpha + b^\alpha\nu^\alpha\, \text{, where}
    \label{eqn:pi}\\
    \tau^\alpha &= \frac{\partial\psi^{\rm{e}}}{\partial\bm{E}^{\rm L}}:(\bm{C}^{\rm
    L}\bm{S}^\alpha) \label{eqn:tau}
\end{align}
is the resolved shear stress on the $\alpha$th slip system, while $b^\phi>0$ and $b^\alpha>0$ denote the inverse mobilities for $\phi$ and $\nu^\alpha$, respectively.\footnote{In \eref{eqn:pi}, we are using the notation $\bm A : \bm B$ to denote the inner product between tensors $\bm A$ and $\bm B$. In indicial notation, $\bm{A}:\bm{B}=A_{ij}B_{ij}$.} To conclude,
equations (\ref{eqn:flow}), (\ref{eqn:u}), (\ref{eqn:phi}) and (\ref{eqn:v}) are the governing
equations for the unknown fields $\bm F^{\rm p}$, $\phi$, $\bm{u}$ and $\nu^\alpha$, respectively.

We note that \erefs{eqn:u}{eqn:tau} are identical to
those derived by \citet{admal2018unified}, wherein the primary unknowns in \eref{eqn:v} are the slip rates $\nu^\alpha$, which have to be solved for each corresponding slip system. This is a major disadvantage, as correctly pointed out by \citet{ask2019cosserat}, in materials with a large number of slip systems --- for example, 
body-centered-cubic materials like $\alpha$-Fe have up to 48 slip systems.
To this end, we note that the slip rate on a specific
slip system is seldom of interest since the motion of GBs, through
the evolution of $\bm{F}^{\rm P}$ via \eref{eqn:flow}, is a combined effect of
slip on all available slip systems. This motivates us to seek an alternative
formulation, which captures the combined effects of all slip rates, while
avoiding the need to solve for each slip rate explicitly. To
    this end, we substitute \eref{eqn:pi} into \eref{eqn:v} and rearrange as
\begin{equation}
    \nu^\alpha = \frac{1}{b^\alpha} \left[ {\rm{Div}} \, \bm{\xi}^\alpha - \frac{\partial\psi^{\rm{gb}}}{\partial\bm{G}}:(\bm{S}^\alpha\bm{G} + \bm{G}(\bm S^\alpha)^T) + \tau^\alpha \right].
    \label{eqn:v_arranged}
\end{equation}
From \eref{eqn:flow}, $\bm{L}^{\rm P}$ can be expressed as
\begin{equation}
    \bm{L}^{\rm P} = \sum_{\alpha=1}^{N_s}\nu^\alpha\bm{S}^\alpha = \dot{\bm{F}^{\rm P}}(\bm{F}^{\rm P})^{-1}.
    \label{eqn:lp}
\end{equation}
Substituting \eref{eqn:v_arranged} into \eref{eqn:lp}, and rearranging, we have
\begin{equation}
    \sum_{\alpha=1}^{N_s} \left[ \frac{\bm{S}^\alpha}{b^\alpha}{\rm{Div}} \, \bm{\xi}^\alpha \right] = \sum_{\alpha=1}^{N_s} \left[ \frac{\bm{S}^\alpha}{b^\alpha} \left( \frac{\partial\psi^{\rm{gb}}}{\partial\bm{G}}:(\bm{S}^\alpha\bm{G} + \bm{G}(\bm{S}^\alpha)^T) - \tau^\alpha \right) \right] + \dot{\bm{F}^{\rm P}}(\bm{F}^{\rm P})^{-1}.
    \label{eqn:fp}
\end{equation}
Before proceeding to simplifying \eref{eqn:fp}, we consider explicit functional forms for the inverse mobilities $B^\alpha$ and $b^\alpha$ appearing in \eref{eqn:xi} and \eref{eqn:pi} respectively.

In this work, since we focus primarily on GB plasticity, we assume inverse mobilities $b^\alpha$s are of the form 
\begin{equation}
    b^\alpha = \frac{\overline{b}(\phi)}{c^\alpha},
    \label{eqn:b_alpha}
\end{equation}
where $\overline{b}$ denotes a nominal inverse mobility, and is defined as
\begin{equation}
        \overline{b} = [ m_{\rm min} + (1 - \phi^3 (10-15\phi+6\phi^2))( m_{\rm max} - m_{\rm min} ) ]^{-1}.
    \label{eqn:mobility}
\end{equation}

The quantities $m_{\rm min}$ and $m_{\rm max}$ are the minimum and maximum mobilities, respectively. In addition, we assume $B^\alpha=c_B b^\alpha$, where the scaling coefficient $c_B$ is a constant with dimension of length. The functional form of $\overline b$ in \eref{eqn:mobility} ensures $\overline b = m_{\rm min}^{-1}$ in the grain interiors where $\phi \approx 1$, and by choosing $m_{\rm min}^{-1} \gg m_{\rm max}^{-1}$, plastic slip in the grain interiors is suppressed. The factor $c^\alpha$ is a dimensionless slip system-dependent availability factor --- $c^\alpha=0$ disables the $\alpha^{th}$ slip system, while $c^\alpha=1$ ensures the slip system is available with inverse mobility $\overline{b}$.

Recalling the definitions of the energetic and dissipative components of $\bm \xi^\alpha$ introduced in \eref{eqn:xi}, we note that except for the dissipative part of the left-hand-side of \eref{eqn:fp}, i.e. 
\begin{equation}
    \sum_{\alpha=1}^{N_s} \left[ \frac{\bm{S}^\alpha}{b^\alpha}{\rm{Div}} \, (B^\alpha \nabla \nu^\alpha) \right],
    \label{eqn:fp_lhs}
\end{equation}
all terms in \eref{eqn:xi} are in terms of $\bm F^{\rm P}$. Therefore, we now draw our attention to (\ref{eqn:fp_lhs}). Using \eref{eqn:lp}, the definition of $B^\alpha$ described above, and noting that $\bm S^\alpha$ is a constant tensor, the expression in (\ref{eqn:fp_lhs}) simplifies as
\begin{equation}
    \sum_{\alpha=1}^{N_s} \left[ \frac{\bm{S}^\alpha}{b^\alpha}{\rm{Div}} \, (B^\alpha\nabla\nu^\alpha) \right] = \frac{c_b}{\overline{b}} \frac{\partial \overline{b}}{\partial \phi} \,\nabla [\dot{\bm{F}^{\rm P}}(\bm{F}^{\rm P})^{-1}] \bullet \nabla \phi + c_b \triangle [\dot{\bm{F}^{\rm P}}(\bm{F}^{\rm P})^{-1}].
    \label{eqn:fp_lhs_expanded}
\end{equation}
where $\nabla \bm A \bullet \bm a$ is a second-order tensor whose $ij$-component is defined as $A_{ij,k} a_k$.\footnote{We have used the identity $\mathrm{Div}(a\bm A) = a \mathrm{Div} \bm A + \bm A \nabla a$ for some scalar field $a$ and tensor field $\bm A$ to arrive at \eref{eqn:fp_lhs_expanded}} Finally, substituting \eref{eqn:fp_lhs_expanded} into \eref{eqn:fp}, we obtain
\begin{equation}
\begin{split}
    \sum_{\alpha=1}^{N_s} \left[ \frac{\bm{S}^\alpha}{b^\alpha}{\rm{Div}} \, \bm{\xi}^\alpha_{\rm e} \right] + \frac{c_b}{\overline{b}} \frac{\partial \overline{b}}{\partial \phi} \nabla [\dot{\bm{F}^{\rm P}}(\bm{F}^{\rm P})^{-1}] \bullet \nabla \phi  + c_b \triangle [\dot{\bm{F}^{\rm P}}(\bm{F}^{\rm P})^{-1}] &= \\ \sum_{\alpha=1}^{N_s} \left[ \frac{\bm{S}^\alpha}{b^\alpha} \left( \frac{\partial\psi^{\rm{gb}}}{\partial\bm{G}}:(\bm{S}^\alpha\bm{G} + \bm{G}(\bm{S}^\alpha)^T) - \tau^\alpha \right) \right] + \dot{\bm{F}^{\rm P}}(\bm{F}^{\rm P})^{-1},
\end{split}
\label{eqn:fp_full}
\end{equation}
which is a system of equations for $\bm{F}^{\rm P}$ that does not involve any slip rates.
The transformation of \eref{eqn:v} into \eref{eqn:fp_full} gives rise to a new formulation wherein the unknown fields are $\phi$, $\bm{u}$ and $\bm{F}^{\rm P}$, and their governing equations are (\ref{eqn:u}), (\ref{eqn:phi}) and (\ref{eqn:fp_full}), respectively. In other words, slip rates are no longer solved for in the new formulation. Compared to \eref{eqn:v}, 
which is a system of $N_s$ equations, \eref{eqn:fp_full} is a system of $d^2$ equations --- where $d$ denotes the space dimension --- our formulation results in a computational advantage whenever $d^2<N_s$. As noted earlier, this is indeed the case for most crystalline materials in three dimensions.
In addition, the boundary conditions (see \eref{eqn:v}) for $\nu^\alpha$, in the formulation by \cite{admal2018unified}, transform into Dirichlet/Neumann boundary conditions for $\bm{F}^{\rm P}$. Finally, we note that although \eref{eqn:v} and \eref{eqn:fp_full} are not equivalent --- satisfying \eref{eqn:fp_full} is a necessary but not sufficient condition for satisfying \eref{eqn:v} on each slip system --- the two equations should, in principle, result in identical evolution for $\bm F^{\rm P}$.

\section{Numerical examples}
\label{sec:numerics}
In this section, four numerical examples are presented to demonstrate different aspects of the model. In the first example, we drive GB migration by the synthetic driving force as introduced in in \sref{sec:energy}. In the second example, a scaling test is presented, where we compare the computational efficiency of the new formulation to the one proposed by \cite{admal2018unified}. The third example concerns the evolution of a tricrystal formed by embedding a circular grain in a bicrystal with a flat GB, a system studied by \citet{trautt2014capillary} using MD simulations. In this example, we examine the evolution of the tricrystal system with and without the application of an external shear stress, and qualitatively compare the results to the MD simulations of \citet{trautt2014capillary}. In the last example, taking advantage of the computational efficiency of our formulations, we simulate GB evolution in a polycrystal. In particular, we  compare the curvature-driven evolution of GBs in a polycrystal under no loads with GB evolution under an external loading condition. This example highlights the key role stress, due to external loads, plays in GB evolution.

All simulations presented in this section were conducted in COMSOL 5.5
\citep{comsol}. The system of equations described in \sref{sec:equations}
are solved in a fully-coupled manner on an Intel Xeon Silver
processor with 40 cores and 2.2 GHz clock speed.

\subsection{Grain boundary motion due to a synthetic potential}
\label{sec:synPot}
In this example, we simulate grain boundary motion induced by an applied synthetic driving force. In addition, we implemented the same synthetic driving force in the formulation by \cite{admal2018unified}, and compare the results of the two formulations.

For the model problem, we consider a bicrystal with a size 20 by 3.33 nm$^2$ centered at $\bm X = (10\,\mathrm{nm}, 10\,\mathrm{nm})$ with a symmetric tilt grain boundary at $X_1=10\, \mathrm{nm}$. The initial condition for the crystal orientation is given by $\bm F^{\rm L}(\bm X,0)=\bm R(\theta^0(\bm X))= (\bm F^{\rm P}(\bm X,0))^T$, where 
\begin{equation}
    \theta^0(x,y) = \frac{\pi}{6} \left[ \frac{1}{2} - \frac{1}{1 - {\rm{exp}}( -6x + 60 )} \right ],
    \label{eqn:initial_theta}
\end{equation}
describes a GB with misorientation of $30^\circ$. The bicrystal
    is fixed on its left end but is free to deform on the right end, yielding a
    free-end condition. Periodic boundary conditions are applied to the top and
    bottom faces of the computational domain to mimic an infinite bicrystal.
    We assume the material is equipped with $12$ slip systems with equal inverse mobilities ($b^\alpha$s). To construct the slip systems, 12 uniformly spaced angles $\beta^\alpha$ in the range $[0,90^{\circ}]$ are generated, and the corresponding slip directions and normals are computed as
\begin{equation}
    \bm{s}^{\alpha} = [ \; {\rm{cos}}(\beta^\alpha) \; , \; {\rm{sin}}(\beta^\alpha) \; ]^T, \;\;
    \bm{m}^{\alpha} = [ \; -{\rm{sin}}(\beta^\alpha) \; , \; \rm{cos}(\beta^\alpha) \; ]^T.
  \label{eqn:slips}
\end{equation}

The GB is driven to the right by adding a synthetic potential of strength $\Delta E=0.147 \, \rm{GPa}$ to the right-hand grain using $f(\phi)$ given in \eref{eqn:modification}. The model parameters used in the simulations discussed in this section are listed in \tref{tab:mat_prop}.
\begin{table}[h]
    \caption{Model parameters used in the simulations in Section \ref{sec:synPot}}
    \centering
    \begin{tabular}{l l l }
    \hline
    $\epsilon^2$ & $8.533\times 10^{-10} \;\; \rm{J}m^{-1}$  \\
    $\alpha^2$   & $1.06\times 10^{-9} \;\; \rm{J}m^{-1}$  \\
    $s$          & $1.7\;\; \rm{J}m^{-2}$  \\
    $e_0$        & $2.1 \;\; \rm{GPa}$  \\
    $\gamma$     & $8.0\times 10^{-7} \;\; $m  \\
    $\lambda$    & $115.7 \;\; \rm{GPa}$  \\
    $\mu$        & $45 \;\; \rm{J}m^{-3}$  \\
    $b^\phi$     & $1.0\times 10^{12} \;\; \rm{kg}/(m \cdot s)$  \\
    $m_{\rm min}$    & $4.0\times 10^{-20} \rm{(m \cdot s)}/kg$ \\
    $m_{\rm max}$    & $4.0\times 10^{-11} \rm{(m \cdot s)}/kg$ \\
    $c_B$   & $1.0\times 10^{-25} \rm{m}$ \\
    $\Delta E^*$ & 0.07 \\
    $c^\alpha$   & 1 \\
    \hline
    \end{tabular}
    \label{tab:mat_prop}
\end{table}

The unknown fields $\phi$, $\bm u$, and $\bm F^{\rm P}$ are interpolated using Lagrange quadratic finite elements.
At this point, it is emphasized that the initial condition for $\bm{F}^{\rm P}$
is a smooth rotation field satisfying the orthogonality condition $(\bm{F}^{\rm
P})^T \bm{F}^{\rm P} = \bm{I}$ everywhere in the domain. Such is not the case if
$\bm{F}^{\rm P}$ is interpolated component-wise using finite element shape
functions. To ensure that the interpolation of initial $\bm{F}^{\rm P}$ is an
orthogonal tensor field, we express $\bm{F}^{\rm P}$ using its right polar
decomposition $\bm{F}^{\rm P} = \bm{R}^{\rm P} \bm{U}^{\rm P}$, and in 2D, the
four components of $\bm{F}^{\rm P}$ are replaced by $\theta^{\rm P}$, $U^{\rm
P}_{11}$, $U^{\rm P}_{12}$ and $U^{\rm P}_{22}$ which are interpolated using
Lagrange quadratic finite elements. This treatment is adopted in \emph{all}
simulations. When solving the equations from both the current
formulation and the framework by \cite{admal2018unified}, a time-dependent
solver with a constant step size of $\Delta t = 1\times10^{-7}$s was used
for time stepping, and both systems were evolved for $5\times10^{-6}$s.

\begin{figure}[h!] 
    \centering
     \subfloat[]{
         \includegraphics[width=0.48\textwidth]{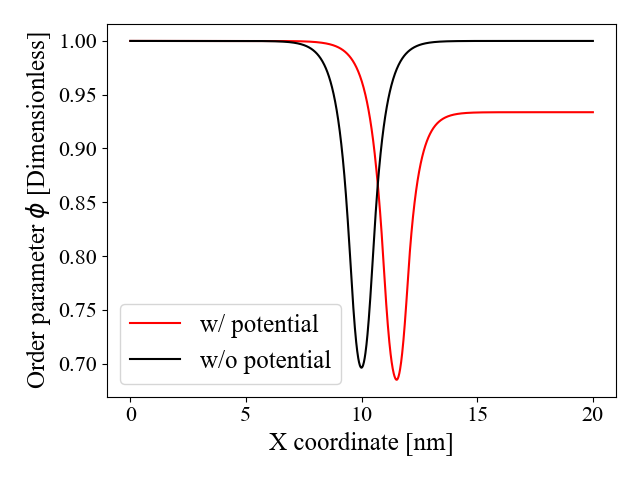}
         \label{fig:synthetic_phi}
     }
     \subfloat[]{
         \includegraphics[width=0.48\textwidth]{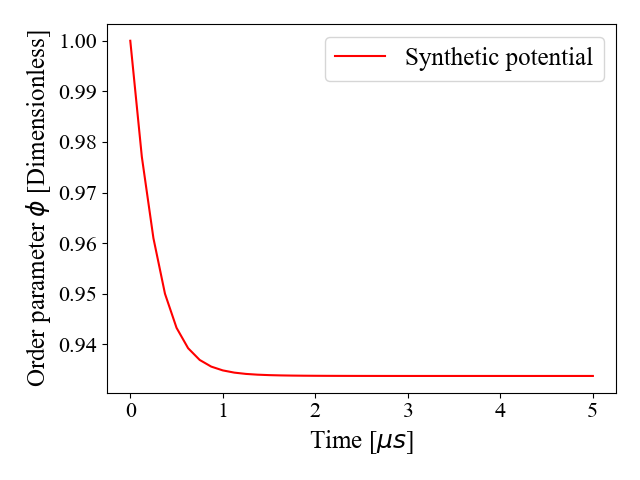}
         \label{fig:phi_his}
     }
     \caption{\psubref{fig:synthetic_phi} A comparison showing
     the effect of the synthetic potential on the steady state of $\phi$. 
    \psubref{fig:phi_his} The evolution of the value of $\phi$ in the
    right grain in the presence of a chemical potential. For the
case with synthetic potential, the minimum of $\phi$ is shifted to the right due to
GB migration. The added potential, which increases the energy of right grain, results in an asymmetry in $\phi$, and $\phi$ decreases gradually and
converges to a steady state value corresponding to a higher energy state.}
    \label{fig:phi_line}
\end{figure}

First, the effect of the synthetic potential on the order parameter $\phi$ is discussed. The contour of $\phi$ at the end of the simulation, extracted from the current framework, is plotted in \fref{fig:synthetic_phi}, along with that obtained from a simulation with no added potential. A time history plot of the value of $\phi$ on the right-hand grain is shown in \fref{fig:phi_his}.
From \fref{fig:synthetic_phi}, we see that minimum of $\phi$ has clearly shifted
to the right, indicating rightward GB migration, as expected. We also notice
that the added potential manifests as an asymmetry in $\phi$ with $\phi<1$ in
the right grain. As the GB migrates, the value of $\phi$ in the
right grain decreases gradually and converges to its steady-state value, as
evidenced in \fref{fig:phi_his}. The lower value of $\phi$ in the right
grain is a consequence of the synthetic potential, interpreted as energy due
to SSDs, added to the right grain. While at the continuum scale SSDs do not
result in lattice distortions, at the atomic scale, they contribute to
disorder. Since $\phi<1$ describes disorder, the prediction that $\phi<1$ in the
right grain is in line with the physical interpretation of $\phi$.
The synthetic potential constructed in this paper allows us to induce migration of GBs surrounding a particular grain, identified by its crystal orientation, and can be used as a surrogate for many forms of driving forces due to energy differences, either coming from differences in SSD densities, stored strain energy, or plastic anisotropy.

\begin{figure}[h!] 
    \centering
     \subfloat[]{
         \includegraphics[width=0.33\textwidth]{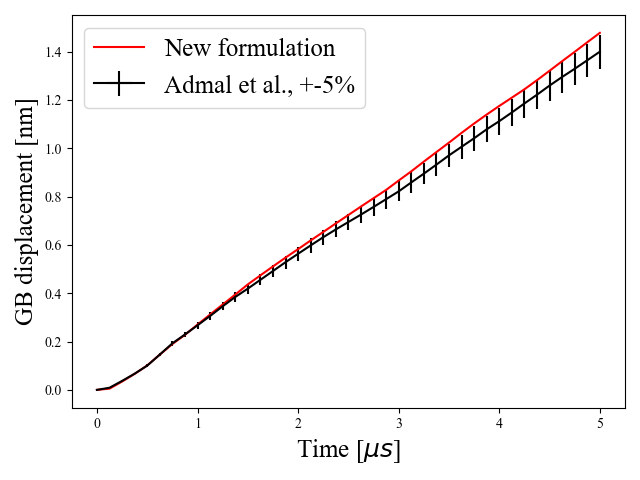}
         \label{fig:disp}
     }
     \subfloat[]{
         \includegraphics[width=0.33\textwidth]{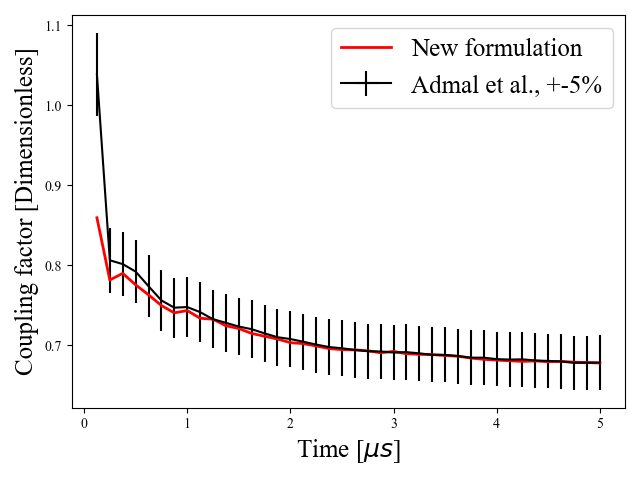}
         \label{fig:cf}
     }
     \subfloat[]{
         \includegraphics[width=0.33\textwidth]{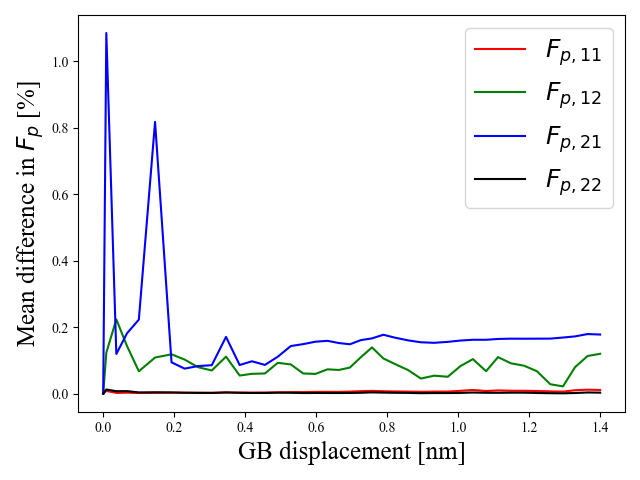}
         \label{fig:fp}
     }
     \caption{Comparison between the current framework and that by \citet{admal2018unified}. \psubref{fig:disp} Time history of the GB displacement, 
    \psubref{fig:cf} time history of the measured coupling factor, and \psubref{fig:fp} mean difference in the components of $\bm F^{\rm P}$. The
    comparison of GB displacements show that the differences are 
    within 5\%. \psubref{fig:cf} and \psubref{fig:fp} show that both frameworks predict nearly identical coupling factors and evolution history for $\bm F^{\rm P}$.}
    \label{fig:accu}
\end{figure}

Next, we leverage this example to compare the results from the
    current formulation to that by \citet{admal2018unified}. The time history
    plots of the GB displacements and measured coupling factor are shown in
    \fref{fig:accu}.
As evidenced by \fref{fig:disp}, some differences in GB displacements are
visible, but they remain below 5\%. The fitted GB migration velocities reveal a
6.07\% difference. However, the simulated coupling factor show good agreement,
as shown in \fref{fig:cf}. These observed differences are within acceptable
ranges, and the difference in migration velocity can be compensated for in model
parameter calibration. In \sref{sec:equations}, it was hypothesized that the
solutions to \eref{eqn:fp_full} and \eref{eqn:v} should yield identical
evolutions for $\bm F^{\rm P}$, we now check the validity of this hypothesis
with simulation data. Recalling that GB plasticity (which is what $\bm F^{\rm
P}$ measures) is due to GB migration, we plot the mean percent difference in
components of $\bm F^{\rm P}$ as a function of the GB displacement, see
\fref{fig:fp}. The plot of the mean differences in components of $\bm F^{\rm P}$
shows good agreement between the evolution histories computed from the two formulations, thus confirming our hypothesis.

\subsection{Performance comparison}
\label{sec:scale}
In this example, we compare the computational cost of the current formulation to that by \cite{admal2018unified} using a similar model problem as discussed in the previous section, where the scaling of total number of DOFs and simulation time are measured. We limit our comparison to the framework by \cite{admal2018unified} since it is closely related to our current work and includes identical physics. It is true that many other models for GB migration exist, but a comparison of numerical efficiency is not fair unless the models being compared include similar physics. To this end, it is appropriate to compare our work with some dislocation-based models such as \cite{zhang2018motion} and \cite{zhang2020new}, and disconnection-based models such as \cite{runnels2020phase} and \cite{gokuli2021multiphase}. However, as those models are still under active development and are not formulated for numerical efficiency, further comparison is not pursued here.

\begin{figure}[h!] 
    \centering
     \subfloat[]{
         \includegraphics[width=0.48\textwidth]{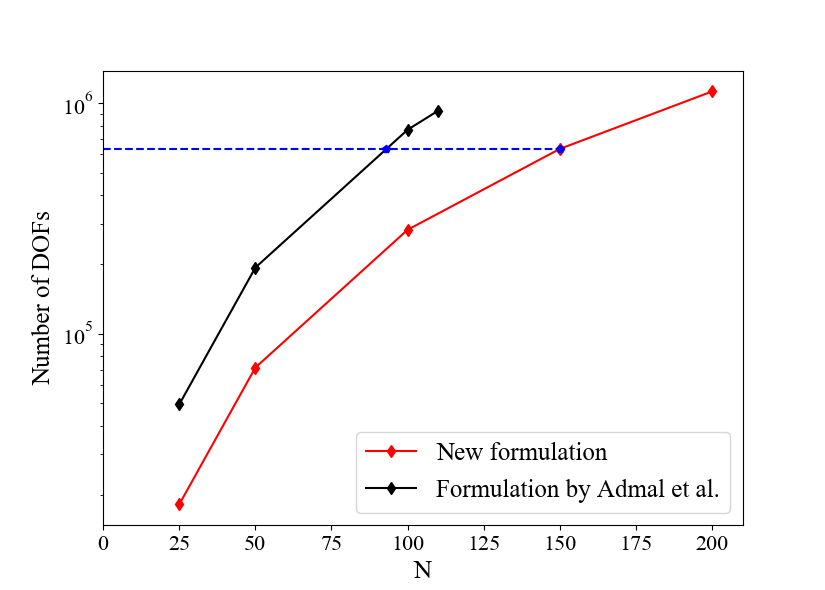}
         \label{fig:n_dof}
     }
     \subfloat[]{
         \includegraphics[width=0.48\textwidth]{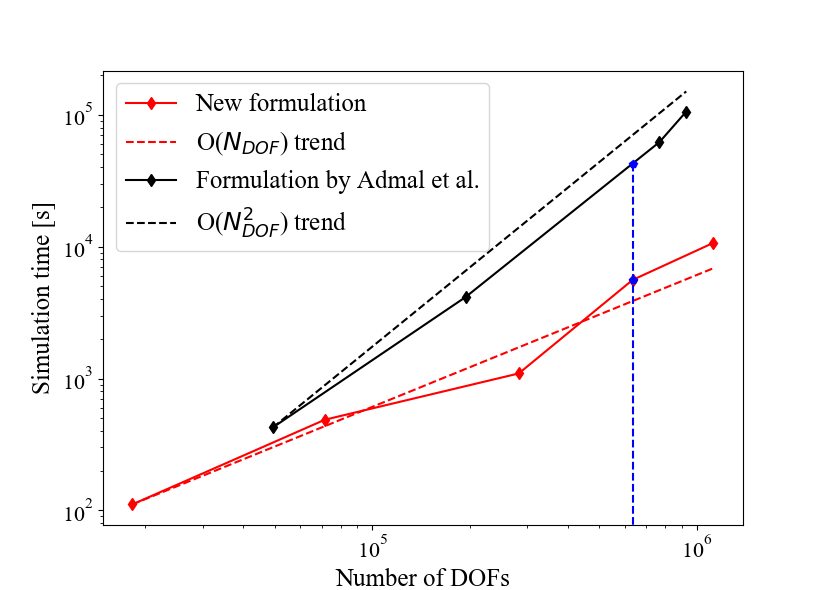}
         \label{fig:dof_t}
     }
    \caption{Scaling tests comparing our formulation with that of \citet{admal2018unified}: \psubref{fig:n_dof} a plot comparing the of number of DOFs versus the mesh size, and
    \psubref{fig:dof_t} simulation time (wall time) as a function of number of DOFs. The points having identical number of DOFs are highlighted in blue for later comparison. The new framework yields far fewer DOFs for a given mesh, which translates to shorter simulation time.}
    \label{fig:scaling}
\end{figure}

In order to explore the scaling of our formulation, we change the size of the bicrystal to a square with an edge length of 20 nm and consider a sequence of uniform meshes with $N^2$ elements, generated using the sequence $N = \{ 25 , 50 , 100 , 110 , 150 , 200 \}$. All other simulation settings are identical to that described in \sref{sec:synPot}. A plot comparing the number of DOFs in each mesh, and the corresponding simulation time is shown in \fref{fig:scaling}.
The reduction in the number of DOFs, and its dependence on $N$ is shown in \fref{fig:n_dof}. In this example, the
framework in \cite{admal2018unified} requires a total of 19 DOFs (1 for $\phi$, 2
for $\bm{u}$, 12 for $\nu_\alpha$ and 4 for $\bm{F}^{\rm P}$) per node. While using the current formulation, the number of DOFs per node is reduced to 7 (no DOFs
related to $\nu_\alpha$), which is a 63.2\% decrease. The reduced number of DOFs
in turns leads to a huge decrease in simulation time, as evidenced in \fref{fig:dof_t}.
Apart from the direct advantage that the new formulation yields fewer number of
DOFs for a given mesh, it is interesting to note that even when the number of
DOFs are identical (e.g., two blue points in \fref{fig:n_dof}), the new
formulation has a shorter run time (e.g., comparing the two blue points in
\fref{fig:dof_t}, shows that the simulation time of the formulation by
\cite{admal2018unified} is about 6.6 times longer). We hypothesize that this
additional gain in performance is due to the change in sparsity pattern (hence
the bandwidth) of the system matrix associated with fewer DOFs per node, as the
solution time for sparse matrix solvers depend both on the system size and its
sparsity pattern. It should be noted that although the scaling test was
performed on a simplified geometry, similar performance enhancements are
expected in general. The reason for this claim is that the performance gain
roots from the reduced number of DOFs per node, which is independent of the
underlying geometry or the polycrystal system. This significant increase in
computational efficiency, resulting from fewer DOFs per node and improved
sparsity pattern, is beneficial in many ways. For more complex crystal
configurations, a refined mesh near the GBs is desired to resolve high local
gradients. The current formulation relaxes the requirements on computational
cost, giving room for the use of a more locally refined mesh. In addition,
reduced number of DOFs entails less memory usage during the simulations.

\subsection{Evolution of a circular tricrystal}
\label{sec:tricrystal}
\begin{figure}[h!] 
    \centering\includegraphics[scale=0.7]{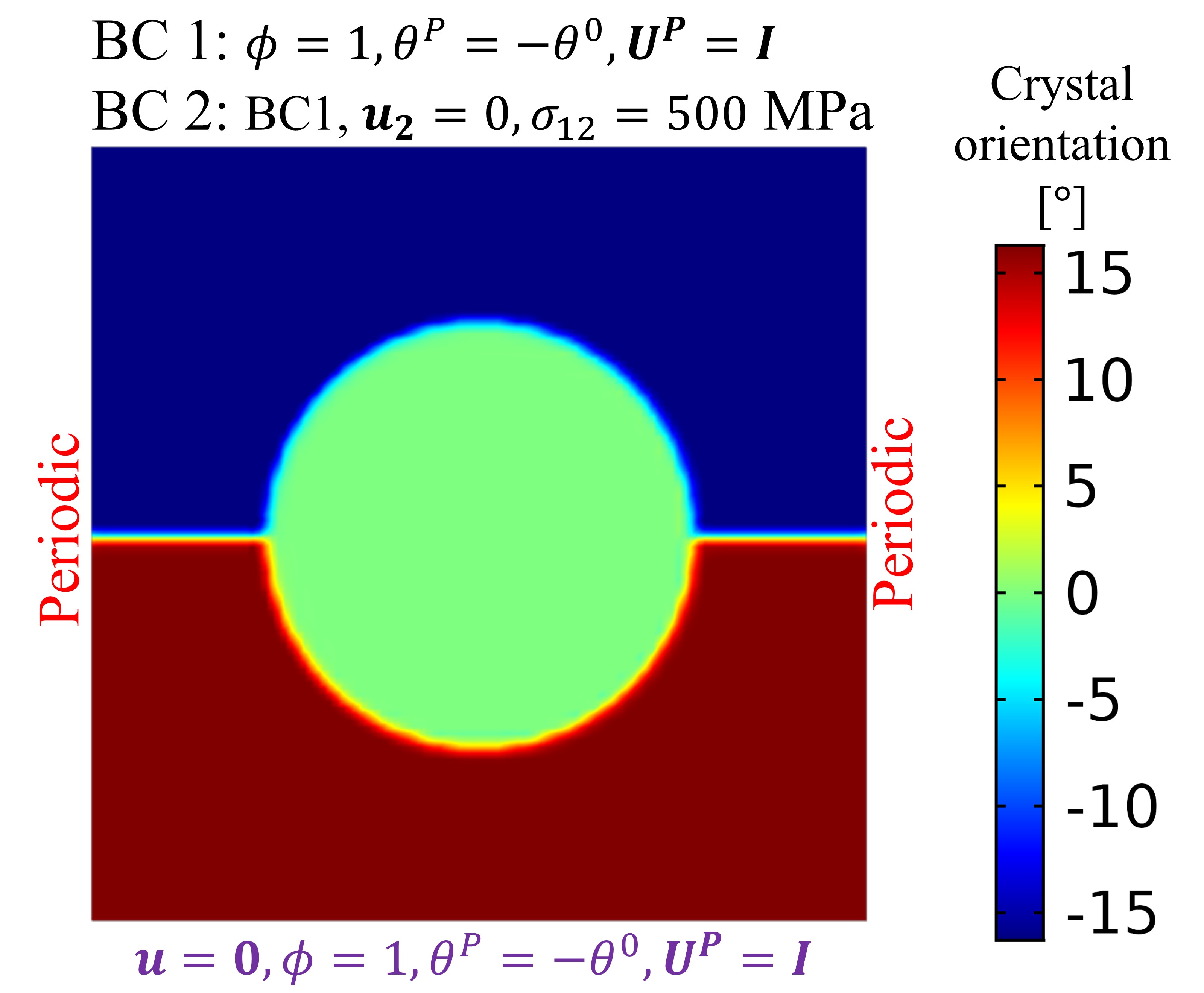}
    \caption{A schematic of the tricrystal along with the imposed boundary
    conditions. Laplacian smoothing of $\theta^0$ is used to generate a smooth initial field. }
    \label{fig:tc}
\end{figure}

In this example, we study the evolution of a copper tricrystal, originally
constructed by \citet{trautt2014capillary}, that is formed by embedding a
circular grain into a bicrystal, as shown in \fref{fig:tc}. In
    particular, we make a quantitative comparison with predictions from MD, and explore the effect of shear stress on the GB motion and grain
    rotation. All results are are compared to the MD simulations by
    \citet{trautt2014capillary}.

The three grains are oriented such that their $[001]$ crystal directions are all aligned along the $z$-axis pointing vertically out of the page. 
A circular grain of radius 8.33 nm with an orientation of 0$^{\circ}$ is placed in the center of a square bicrystal of side length 30 nm, whose upper and lower grains have orientations of 73.74$^{\circ}$ and 16.26$^{\circ}$, respectively. The $[1 0 0]$ crystal direction for the center grain is parallel to the $x$-axis. The computational domain is discretized with quadratic Lagrange elements. This configuration corresponds to Case 1 in the MD study by \cite{trautt2014capillary}. We note that due to the four-fold symmetry of the cubic crystal, the orientations of the grains are not uniquely defined, and this manifests as non-uniqueness of the GND density. For example, the orientations described above result in a larger GND density for the upper semi-circular grain boundary relative to its lower counterpart, while changing the orientation of the top grain to an equivalent $-16.26^\circ$ results in equal misorientations and GND densities. Since the two semi-circular GBs have identical atomistic structures, and therefore equal energies, we proceed with the latter choice. The tricrystal is equipped with three slip systems with equal inverse mobilities, and oriented along directions given by \eref{eqn:slips} with $\beta^\alpha = 0^{\circ}$, $45^{\circ}$, and $-45^{\circ}$. 
To match the initial grain shrinkage rate as observed in MD by
    \cite{trautt2014capillary} (cf. Figure 8b therein), the inverse mobility for
    $\phi$, $b^\phi$, and the maximum slip mobility, $m_{\rm max}$, were
    calibrated as $b^\phi=4.165\times 10^{8} \;\; \rm{kg}/(m \cdot s)$ and
    $m_{\rm max}=1.2\times 10^{-7} \rm{(m \cdot s)}/kg$, respectively. All other
    simulation parameters are identical to those listed in \tref{tab:mat_prop},
except that no synthetic potential is applied.

\begin{figure}[h!]
    \centering
    \begin{tabular}{ c  c  c  c  c  }
    \begin{minipage}[c]{0.21\textwidth}
       \centering 
        \includegraphics[width=\textwidth]{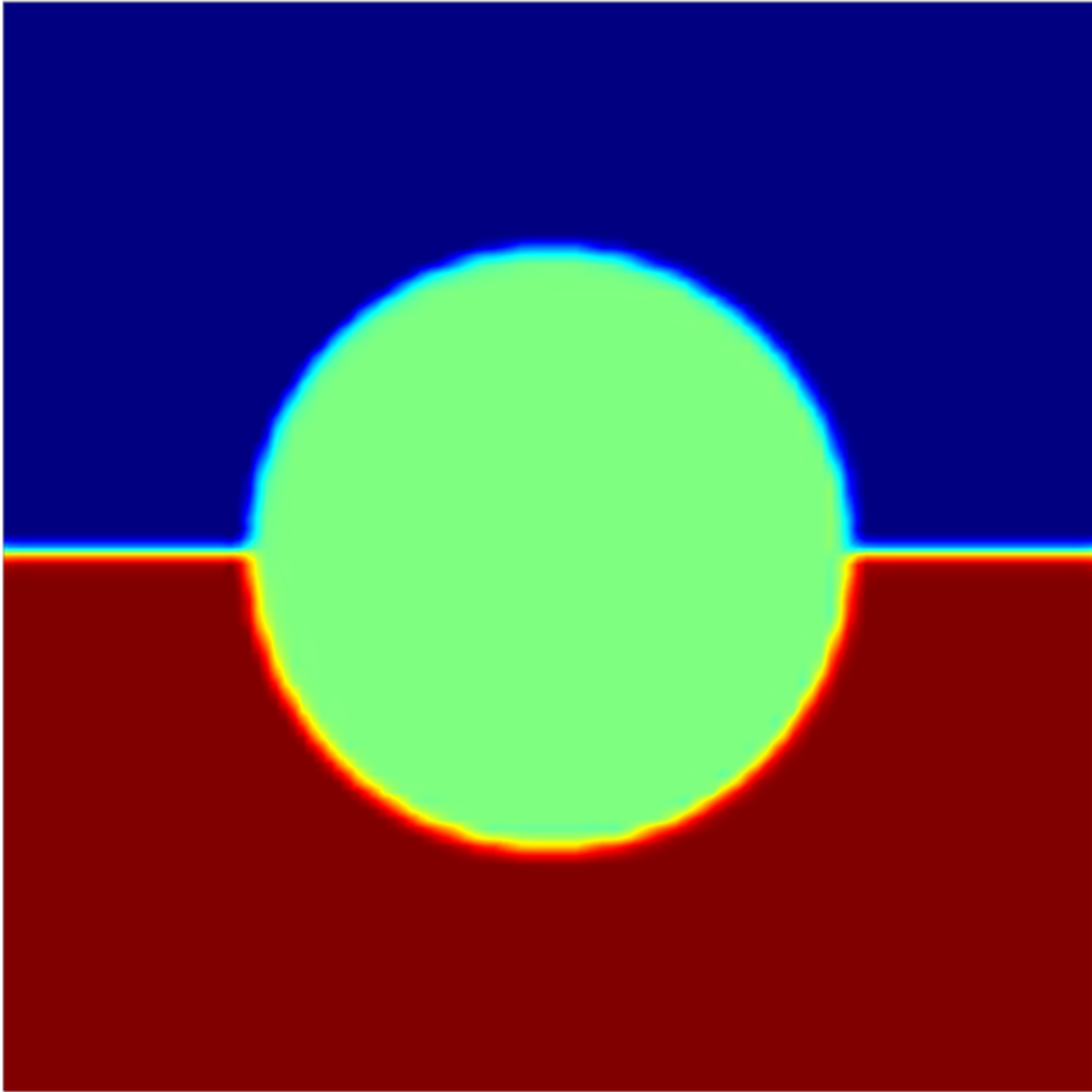} 
        \label{fig:ns_a}
    \end{minipage}
    &
    \begin{minipage}[c]{0.21\textwidth}
       \centering 
        \includegraphics[width=\textwidth]{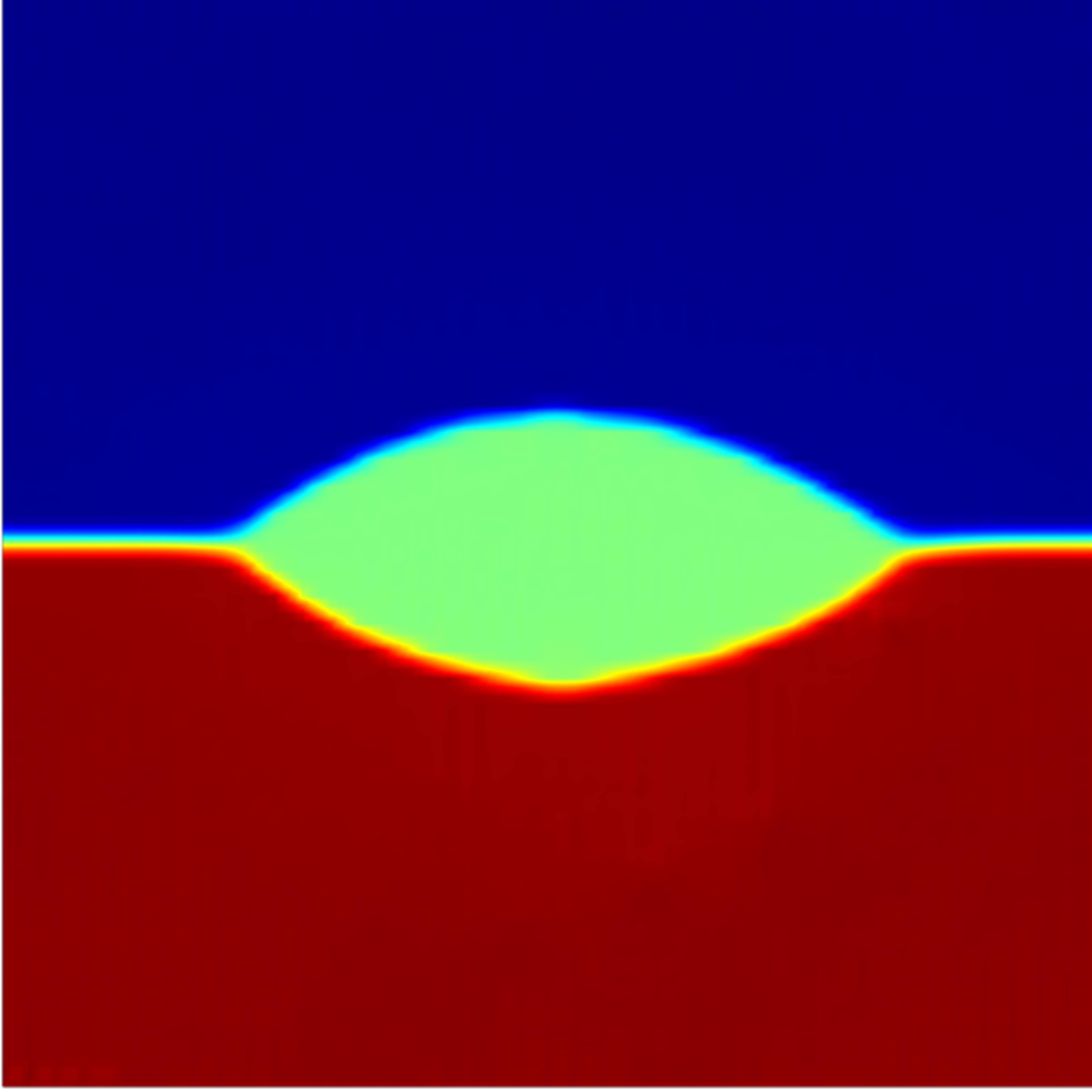}
        \label{fig:ns_b}
    \end{minipage}
    &
    \begin{minipage}[c]{0.21\textwidth}
       \centering 
        \includegraphics[width=\textwidth]{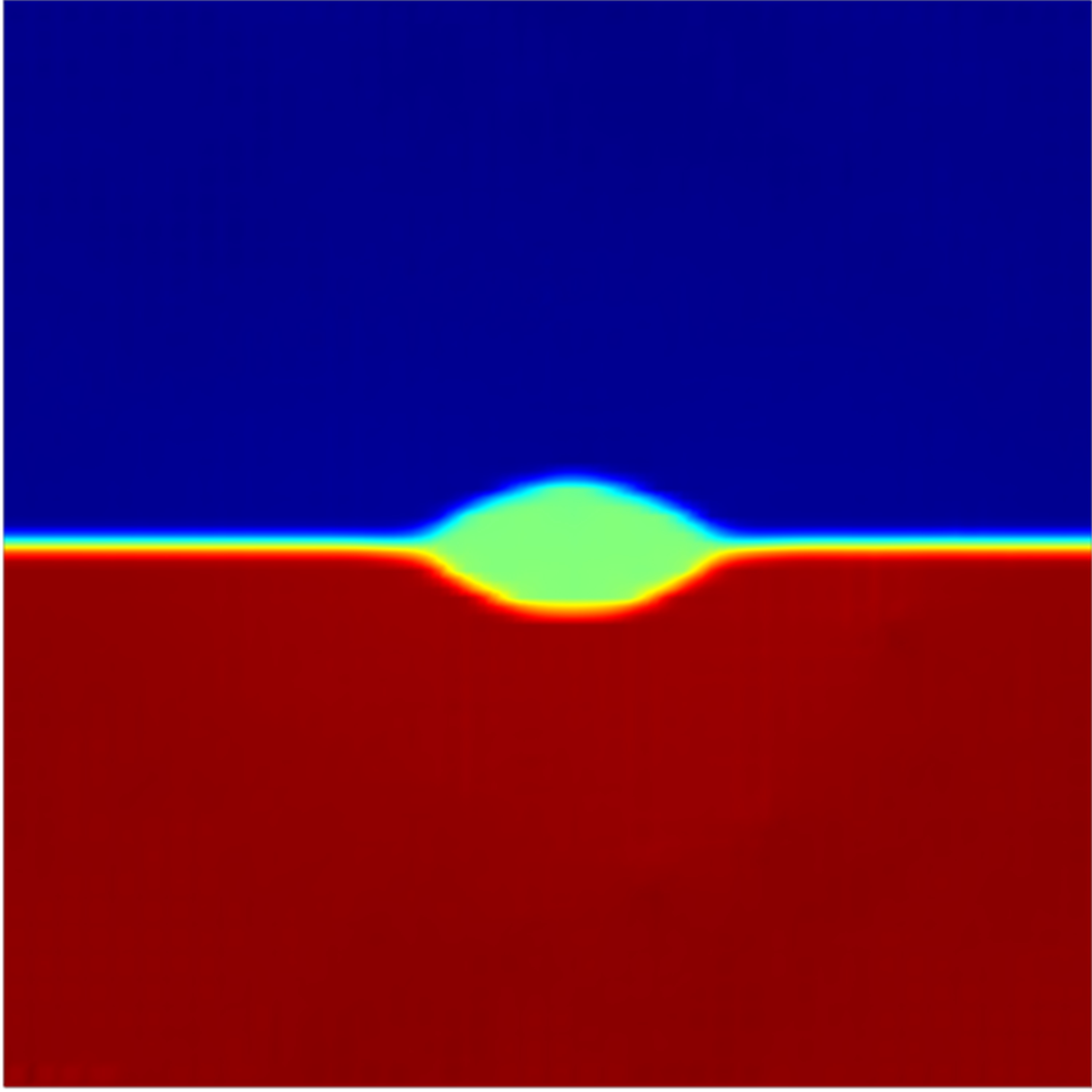}
        \label{fig:ns_c}
    \end{minipage}
    &
    \begin{minipage}[c]{0.21\textwidth}
       \centering 
        \includegraphics[width=\textwidth]{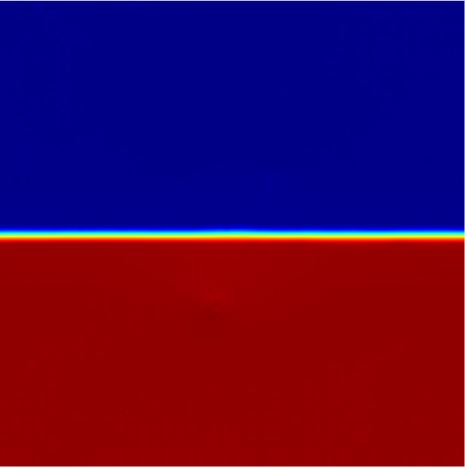}
        \label{fig:ns_d}
    \end{minipage}
    & 
    \begin{minipage}[c]{0.05\textwidth}
       \centering 
        \includegraphics[width=\textwidth]{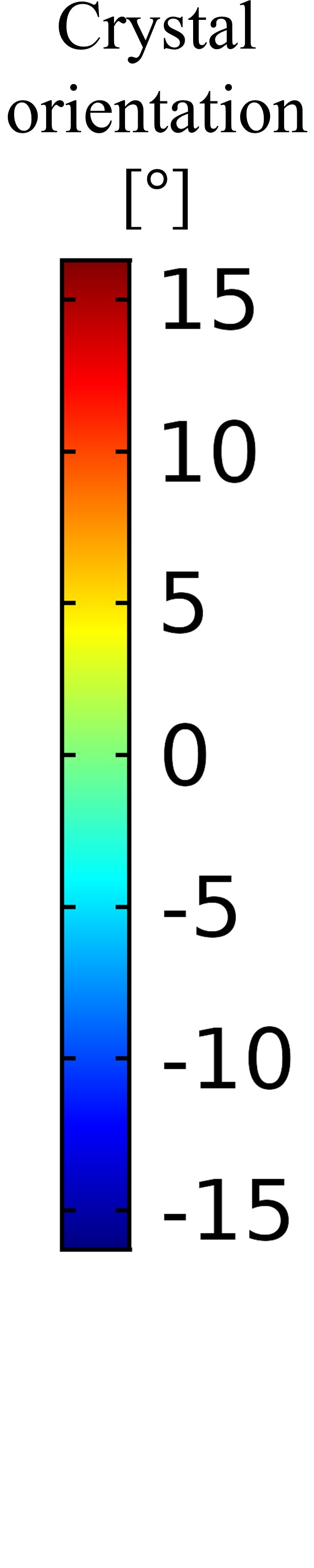}
    \end{minipage}
    \\
    \end{tabular}
    \caption{GB motion in the absence of an external load. Since the top and bottom semi-circular GBs have equal mobilities and opposite shear-coupling factors, symmetry of the GB configuration is preserved, and the center grain shrinks without undergoing any rotation.}
    \label{fig:evo1}
\end{figure}

\begin{figure}[h!] 
    \centering
     \subfloat[]{
         \includegraphics[width=0.5\textwidth]{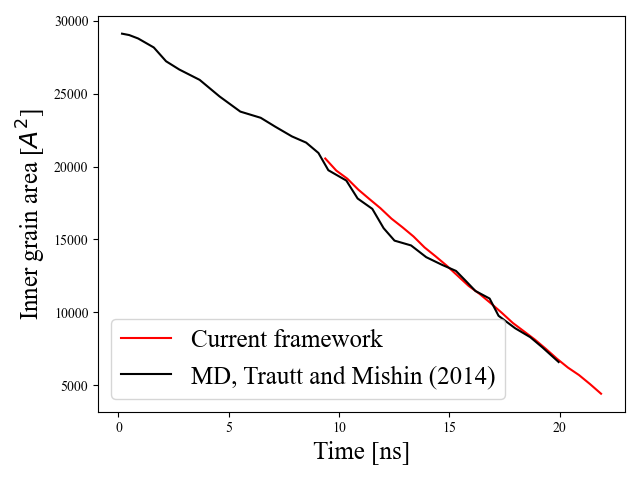}
         \label{fig:area_his}
     }
     \subfloat[]{
         \includegraphics[width=0.5\textwidth]{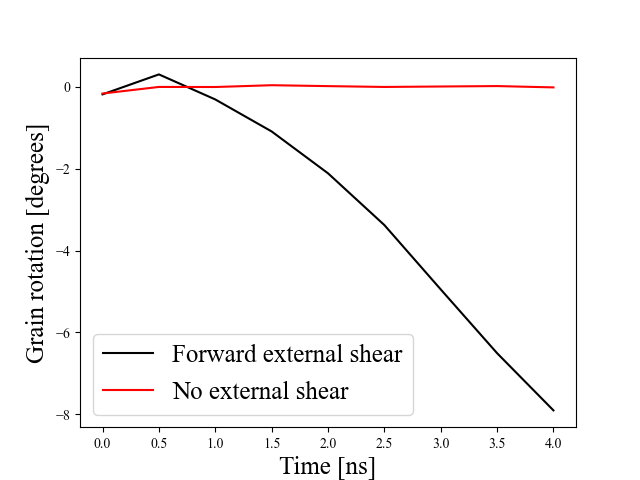}
         \label{fig:rot_his}
     }
     \caption{Time history plots for the tricrystal cases:
         \psubref{fig:area_his} inner grain area as a function of time, where
         our simulation data was shifted to the right by 9.37 ns to match the
         starting grain area, \psubref{fig:rot_his} plot of grain rotation
     versus time when shear stress is applied. In the first case, excellent
 agreement between our simulation and MD data is observed. For the second case,
 the grain undergoes a negative rotation when shear is applied, which is in
 qualitative agreement with MD data. Note that since the applied shear stress
 includes a positive rigid body rotation (as opposed to pure shear), the grain
 orientation angle initially increases.}
    \label{fig:his}
\end{figure}

\begin{figure}[h!]
    \centering
    \begin{tabular}{ c  c  c c c }
    \begin{minipage}[c]{0.21\textwidth}
       \centering 
        \includegraphics[width=\textwidth]{s_a.JPG} 
        \label{fig:wl_a}
    \end{minipage}
    &
    \begin{minipage}[c]{0.21\textwidth}
       \centering 
        \includegraphics[width=\textwidth]{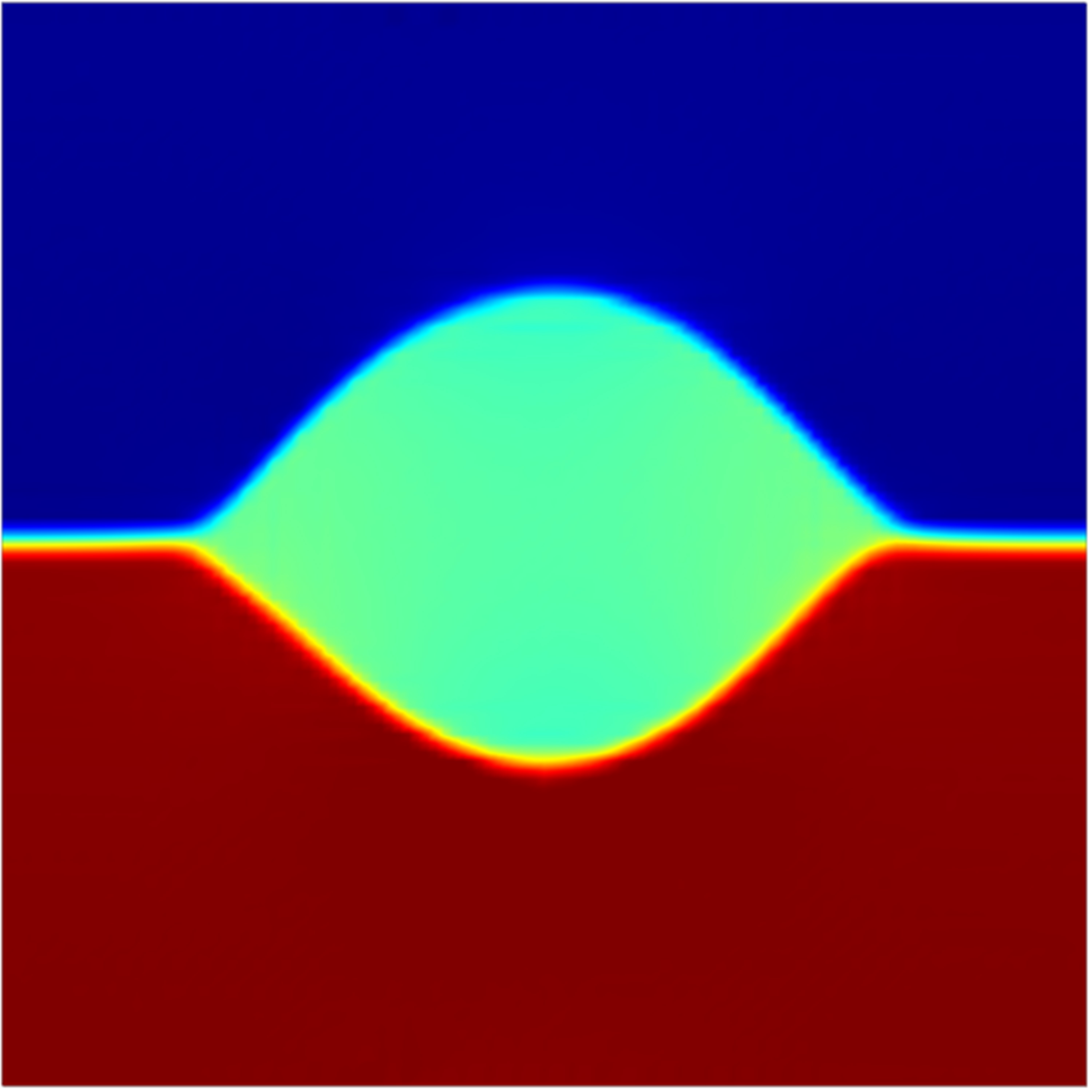}
        \label{fig:wl_b}
    \end{minipage}
    &
    \begin{minipage}[c]{0.21\textwidth}
       \centering 
        \includegraphics[width=\textwidth]{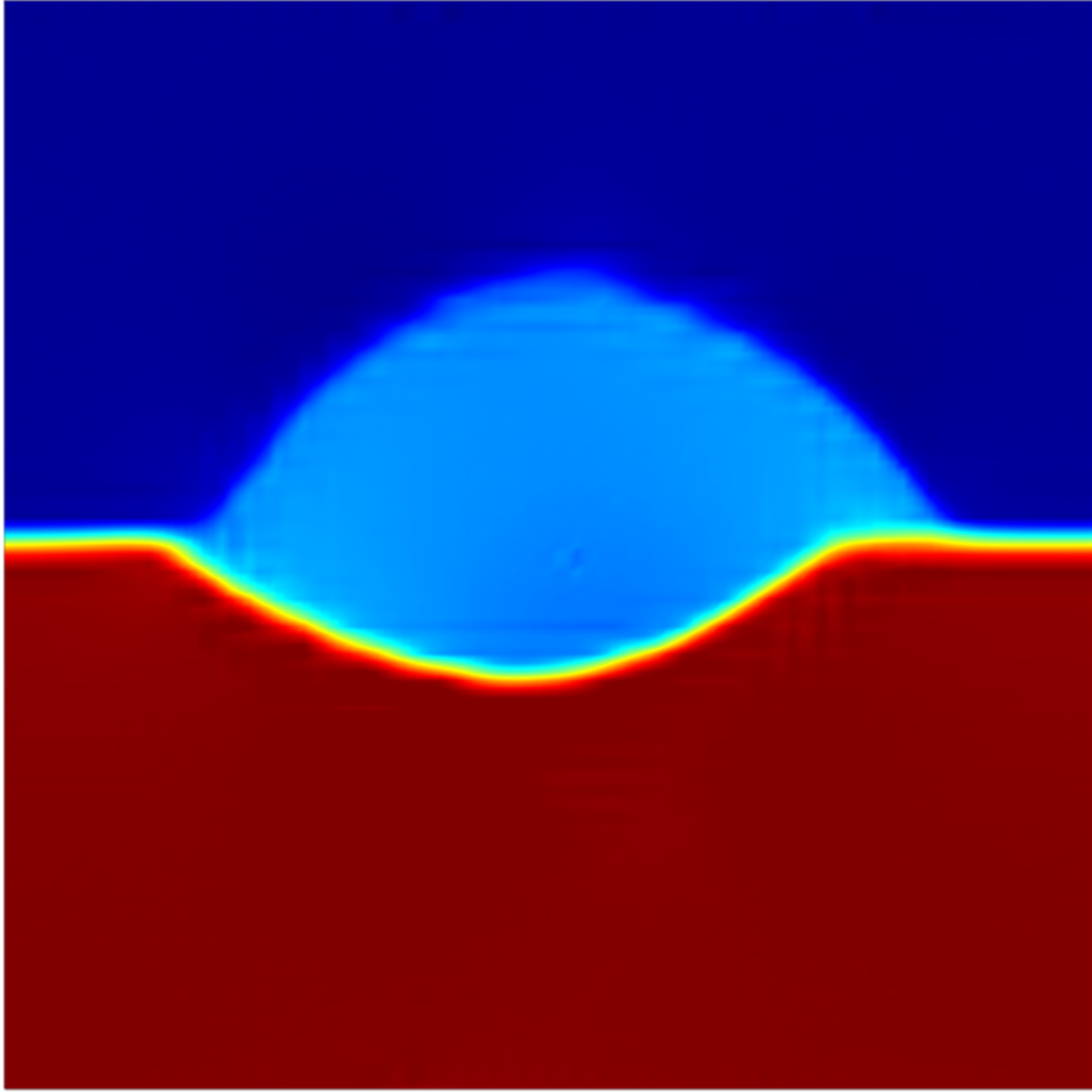}
        \label{fig:wl_c}
    \end{minipage}
    &
    \begin{minipage}[c]{0.21\textwidth}
       \centering 
        \includegraphics[width=\textwidth]{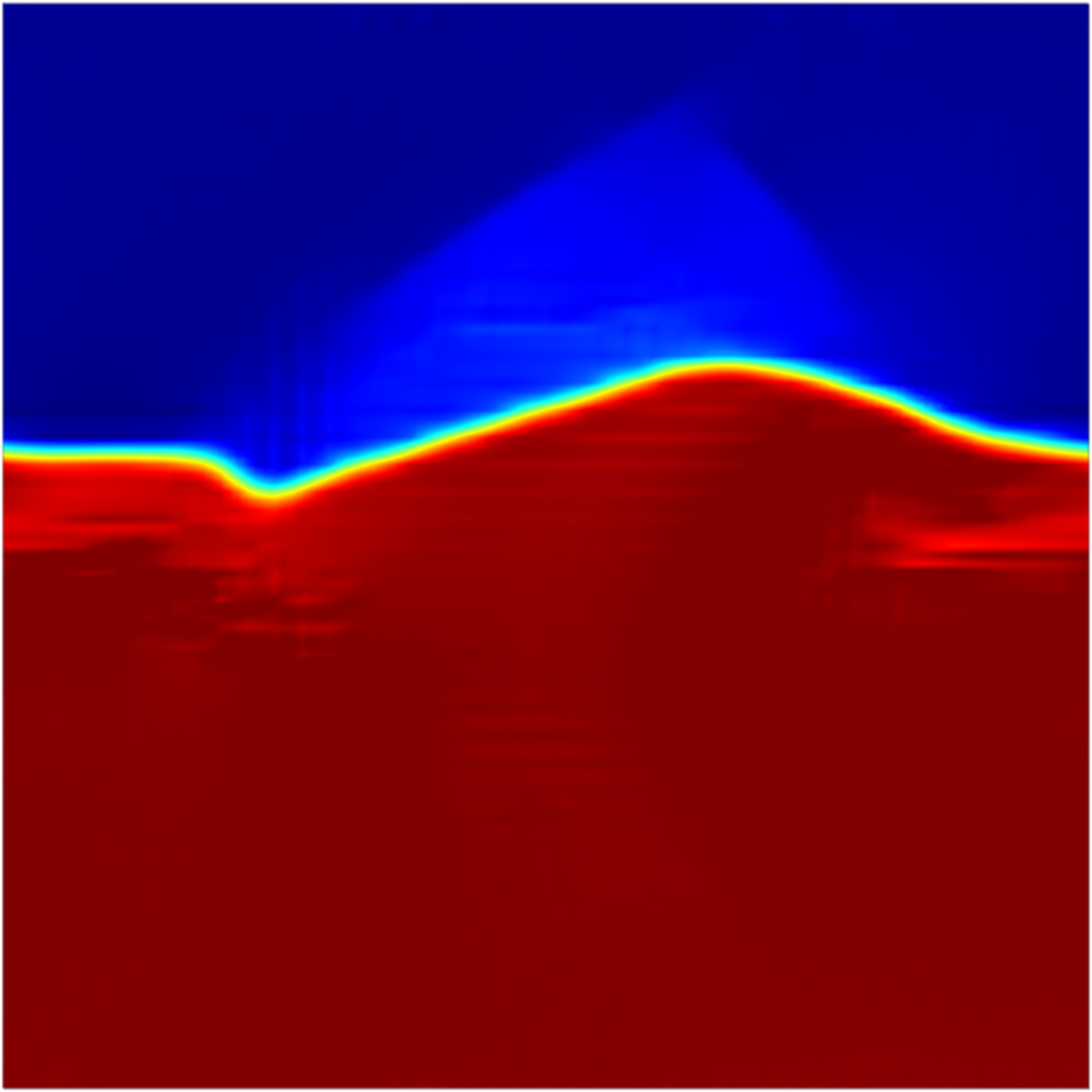}
        \label{fig:wl_d}
    \end{minipage}
    & 
    \begin{minipage}[c]{0.05\textwidth}
       \centering 
        \includegraphics[width=\textwidth]{ColorBar_V2.jpg}
    \end{minipage}\\
    \end{tabular}
    \caption{Evolution of the tricrystal in the presence of an external load. The applied shear stress causes the planar GB to migrate upward, and the collective migration of the two semi-circular GBs induces a negative rotation in the center grain, causing it to decrease in crystal orientation.}
    \label{fig:evo2}
\end{figure}

A schematic showing the initial crystal orientation and the applied boundary
conditions is shown in \fref{fig:tc}. Two cases are considered,
    one with no external loading, and the other with an external shear of 500
    MPa acting on the top surface, and the system is evolved for 19 ns and 14
ns, respectively. Selected snapshots of crystal orientations during their evolution for the two load cases are shown in \fref{fig:evo1} and \fref{fig:evo2}, where the latter corresponds to the case of applied shear. \fref{fig:his} compares the variations of grain area and rotations as functions of time for the two load cases.

From \fref{fig:evo1}, we note that in the absence of external loads the two
semi-circular GBs, driven by curvature, evolve at identical speeds preserving
the symmetry of the grain microstructure. Moreover, the orientation of the
grains are preserved throughout the simulation as shown in \fref{fig:rot_his}.
As noted in \sref{sec:energy}, the GB motion observed in this simulation may be
interpreted as a superposition of coupled GB motion and GB sliding that produces
no grain rotation. These features, which are also observed in MD simulations of
\citet{trautt2014capillary}, can be attributed to coupling factors that are
equal in magnitude but opposite in sign for the two GBs, resulting in no grain
rotation. In the absence of grain rotation, the equivalence (energetic and
kinetic) of the two GBs is preserved, and they continue to evolve with equal
speeds. In addition, the evolution history predicted by the
    continuum simulation is in close agreement with that observed by MD
    simulation, which proves that the framework can produce accurate results
after material parameter calibration.

In the case of shear-driven tricrystal, the GBs experience a driving force due to shear stress in addition to the force due to curvature. \fref{fig:evo2} shows that the planar GB migrates upward due to the external driving force. For the lower semi-circular GB, both the curvature and shear stress drive the GB upwards, whereas for the upper semi-circular GB, the two driving forces have a tendency to drive the GB in different directions (curvature flow drives the GB downwards, while shear stress drives it upwards). This leads to a difference in migration velocities, with the lower GB migrating faster than the upper one. This difference in migration velocities, along with GB coupling, leads to a nonzero net rotation rate for the center grain, in the direction such that its crystal orientation is decreased. When comparing the continuum simulation results with those by MD, we see that the two agree qualitatively, i.e., the planar GB migrates upward, the lower GB migrates at a faster rate than the upper one, and the inner grain decreases in orientation.

The qualitative agreements between continuum and MD simulations discussed above provide confidence for the model to be used in more complex scenarios. While using GB energies and mobilities obtained from atomistics would yield a quantitative comparison with MD simulations, we do not pursue such a study due to certain limitations of our framework, which we will now discuss. In our framework, since the construction of GB energy is in terms of the GND density, and the norm of the GND density increases monotonically with GB misorientation, our model is only applicable for small angle GBs. In the tricrystal example, although the misorientations computed using the originally prescribed orientations ($0^\circ$, $16.26^\circ$ and $73.74^\circ$) are large angles, the cubic symmetry enabled us to identify them with small angles by noting the equivalence of grain orientations $\theta$ and $\theta \pm 90^\circ$, resulting in the correct GND density that accurately reflects the GB energy. While such an unambiguous choice of GND density is possible for small angle GBs, it is not possible for large angle GBs. For example, consider a tricrystal with grain orientations $0^\circ$, $30^\circ$ and $60^\circ$. Using cubic symmetry, we note that all GBs are equivalent with misorientation angle of $30^\circ$. On the other hand, no choice of orientations among their equivalent classes will result in equal misorientations for the three GBs. Therefore, our framework limits us to small angle GBs with a Read--Shockley type GB energy. In \sref{sec:conclusion}, we discuss how our dislocation-based GB model can complement a disconnection-based framework to model GBs with arbitrary misorientations.
 
Finally, we remark that the mobility of a GB depends on $\phi$ through the
relation given in \eref{eqn:b_alpha}, which results in the qualitatively correct
trend, where GBs with larger misorientation have higher mobilities
\cite{trautt2014capillary}.\footnote{MD simulations from
    \cite{trautt2014capillary} highlight the importance of
misorientation-dependent GB mobilities in the shape evolution of the inner
grain, especially when the two semi-circular GBs have vastly different
misorientations.} However, we note that the mobility of the GB also depends on
orientations of the slip systems relative to the GB. Therefore,
    in order to make detailed quantitative comparison between our model and MD,
    further material model parameters (e.g., $b^\phi$, $m_{\rm min}$, $m_{\rm
    max}$ and the functional form of $m$ as a function of $\phi$) calibration is
necessary.

\subsection{Evolution of a polycrystal}
\label{sec:polycrystal}
In this last simulation, we study the evolution of a square polycrystal under two different loading cases. In the first case, no external load is applied. In the second case, a displacement boundary condition is imposed to intentionally induce grain rotation. The evolution of texture in the two cases are compared, and the importance of external loads in GB evolution is highlighted.

The computational domain is a square polycrystal of side length 100 nm with 13 grains, generated using DREAM.3D \citep{groeber2014dream}.\footnote{DREAM.3D is an architecture for computational microstructure tools that allows users to create 'recipes' or pipelines for processing digital instances of microstructure.} The domain is discretized by quadratic Lagrange elements with a uniform mesh size of 1.6 nm. The crystal orientations are drawn at random from a normal distribution with a mean of 22.5$^{\circ}$ and a standard deviation of 10.5$^{\circ}$. Additional checks are performed to limit the GB misorientations to the range 5$^{\circ}$ to 20$^{\circ}$, which ensures that all GBs are within the small angle misorientation range to which our model is applicable. The global X- and Y-axes are assumed to coincide with the $[\overline{1} 1 0]$ and $[0 0 1]$ directions of a single face-centered cubic (fcc) crystal, such that the in-plane effective slip directions are given by \eref{eqn:slips} using $\beta^\alpha = 0^{\circ}$, $54.7^{\circ}$, and $125.3^{\circ}$ \citep{kysar2010experimental}. Other simulation parameters are identical to those listed in \tref{tab:mat_prop}, except that no synthetic potential is applied.

\begin{figure}[h!] 
    \centering
     \subfloat[]{
         \includegraphics[width=0.42\textwidth]{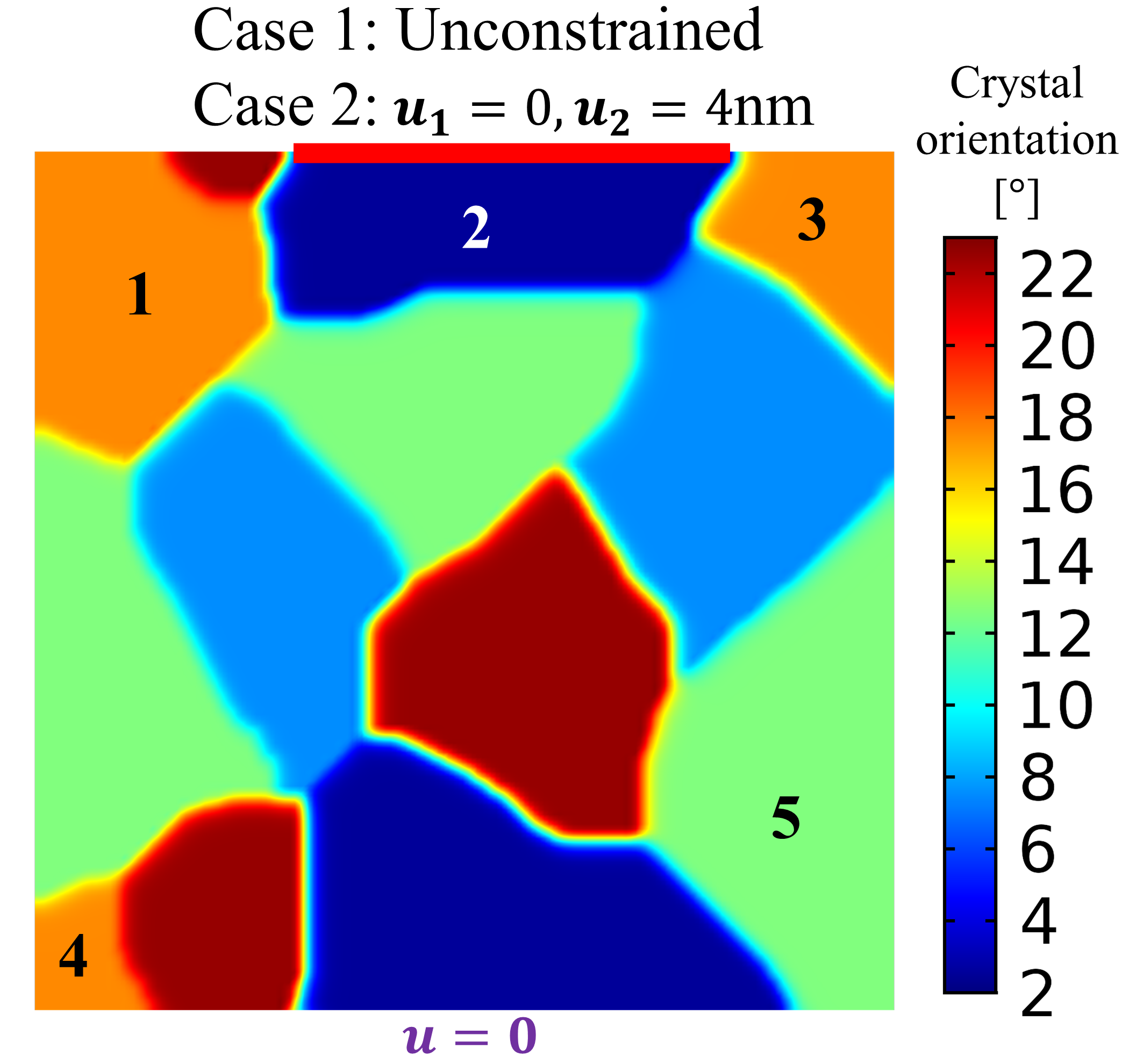}
         \label{fig:poly_ini}
     }
     \subfloat[]{
         \includegraphics[width=0.42\textwidth]{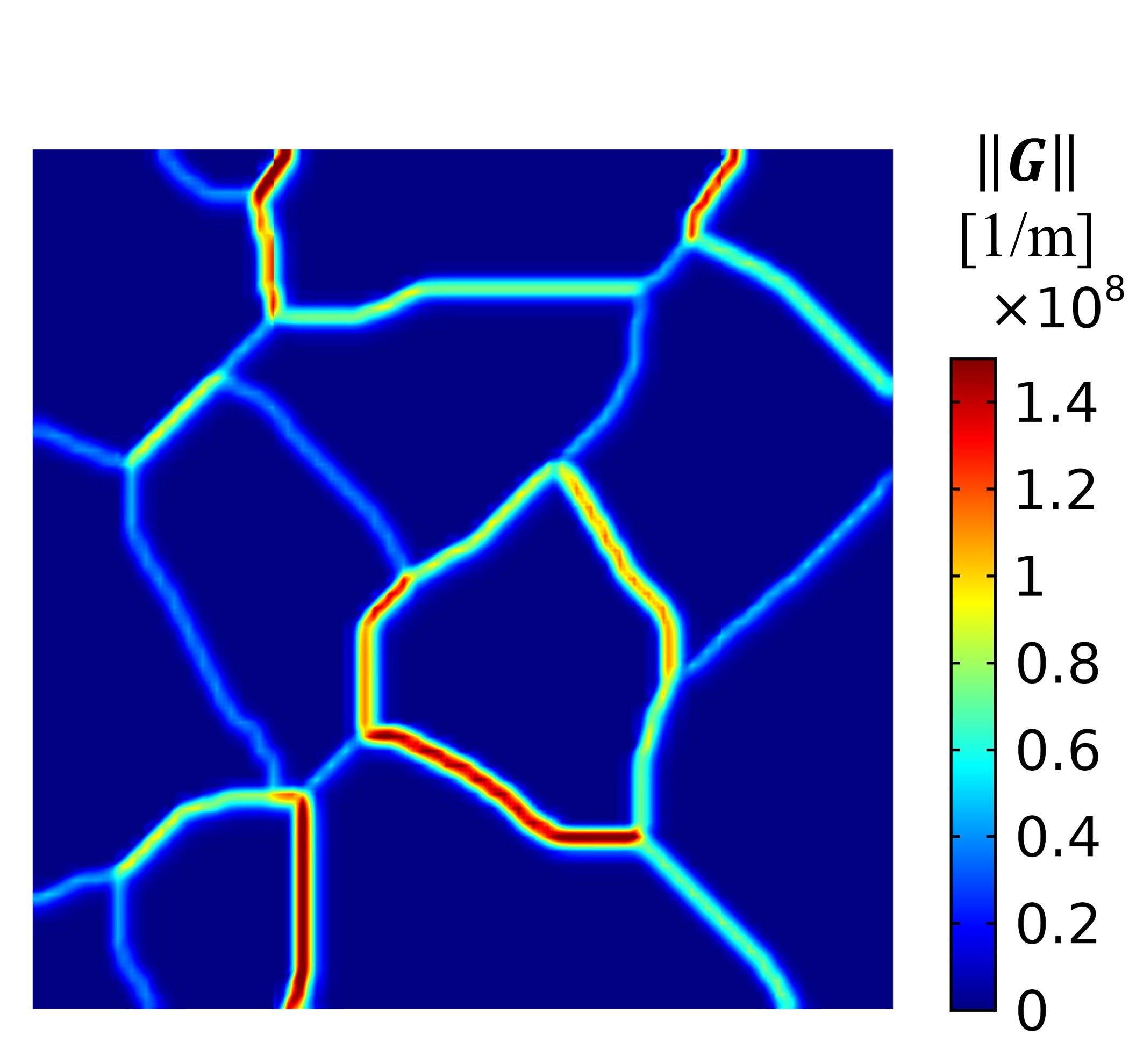}
         \label{fig:poly_ini_GND}
     }
    \caption{\psubref{fig:poly_ini} A color density plot of initial grain orientations of the polycrystal under boundary conditions corresponding to the two simulated load cases. The red line on the top edge corresponds to the region on which a Dirichlet boundary condition is imposed in load case 2. Four grains are labeled for later discussion. \psubref{fig:poly_ini_GND} A color density plot of the GND density corresponding to the initial grain orientations.}
    \label{fig:poly_IC}
\end{figure}

A schematic of the polycrystal with a color density plot of initial grain orientations along with applied boundary conditions for the two loading cases is shown in \fref{fig:poly_ini}. The GND density corresponding to the initial grain orientation is shown in \fref{fig:poly_ini_GND}. Since all GBs in the system are low-angle GBs, the norm of the GND tensor serves as a valid measure of GB misorientation. The two loading cases are simulated using boundary conditions imposed on the top edge of the polycrystal. In the first case, the top edge is traction-free, while in the second case, a vertical displacement of 4 nm is imposed on a part of the boundary, shown in red in \fref{fig:poly_ini}, at a rate of $4\times10^{4}$ nm/s. In both cases, the bottom edge of the domain is held fixed. The duration of the simulation is set to be $2\times10^{-4}$ s in both cases.

\begin{figure}[h!]
    \centering
    \begin{tabular}{ c  c  c  c }
     & & & 
    \multirow{2}{*}{
    \begin{minipage}[c]{0.08\textwidth}
       \centering 
        \includegraphics[width=\textwidth]{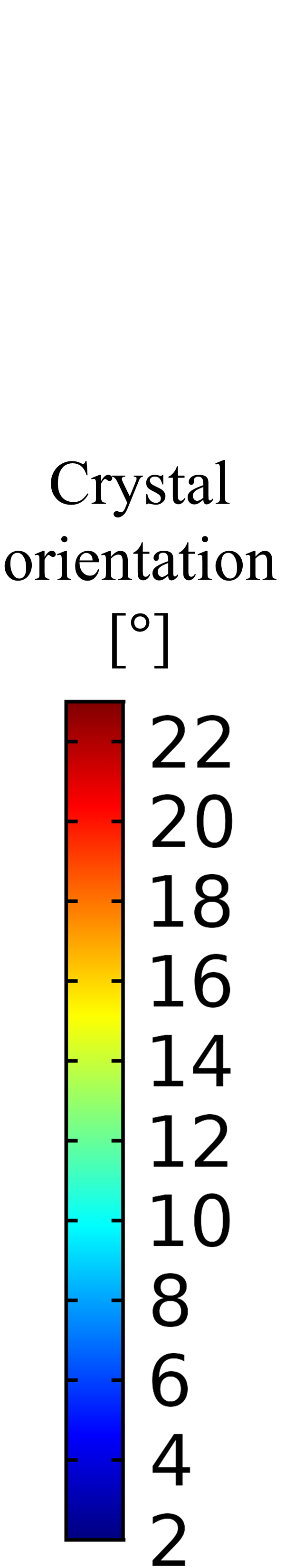}
    \end{minipage}
    }\\
    \begin{minipage}[c]{0.28\textwidth}
       \centering 
        \subfloat[$t=0$]{\includegraphics[width=\textwidth]{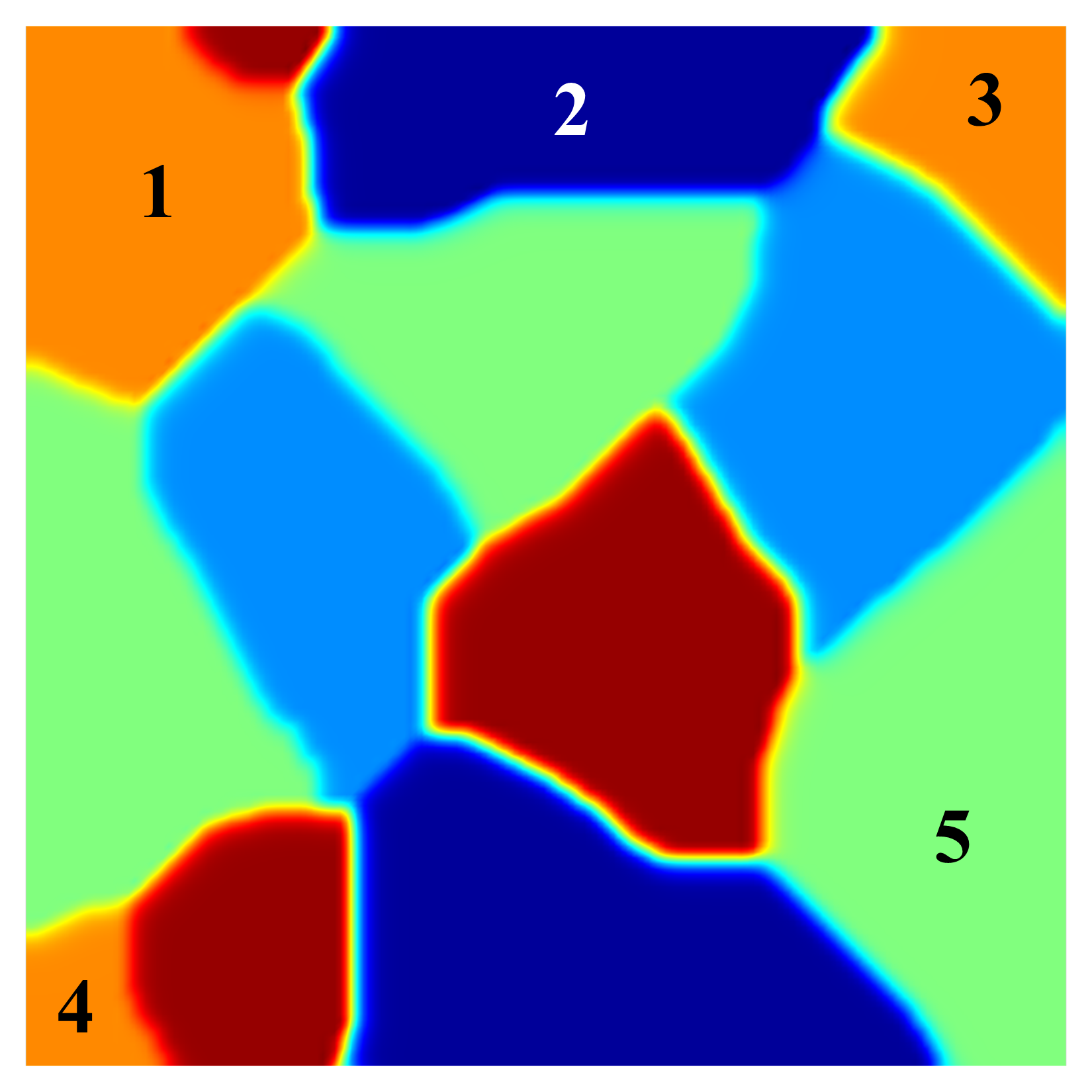}
        \label{fig:te_nl_1}}
    \end{minipage}
    &
    \begin{minipage}[c]{0.28\textwidth}
       \centering 
        \subfloat[$t=1\times10^{-5}$s]{\includegraphics[width=\textwidth]{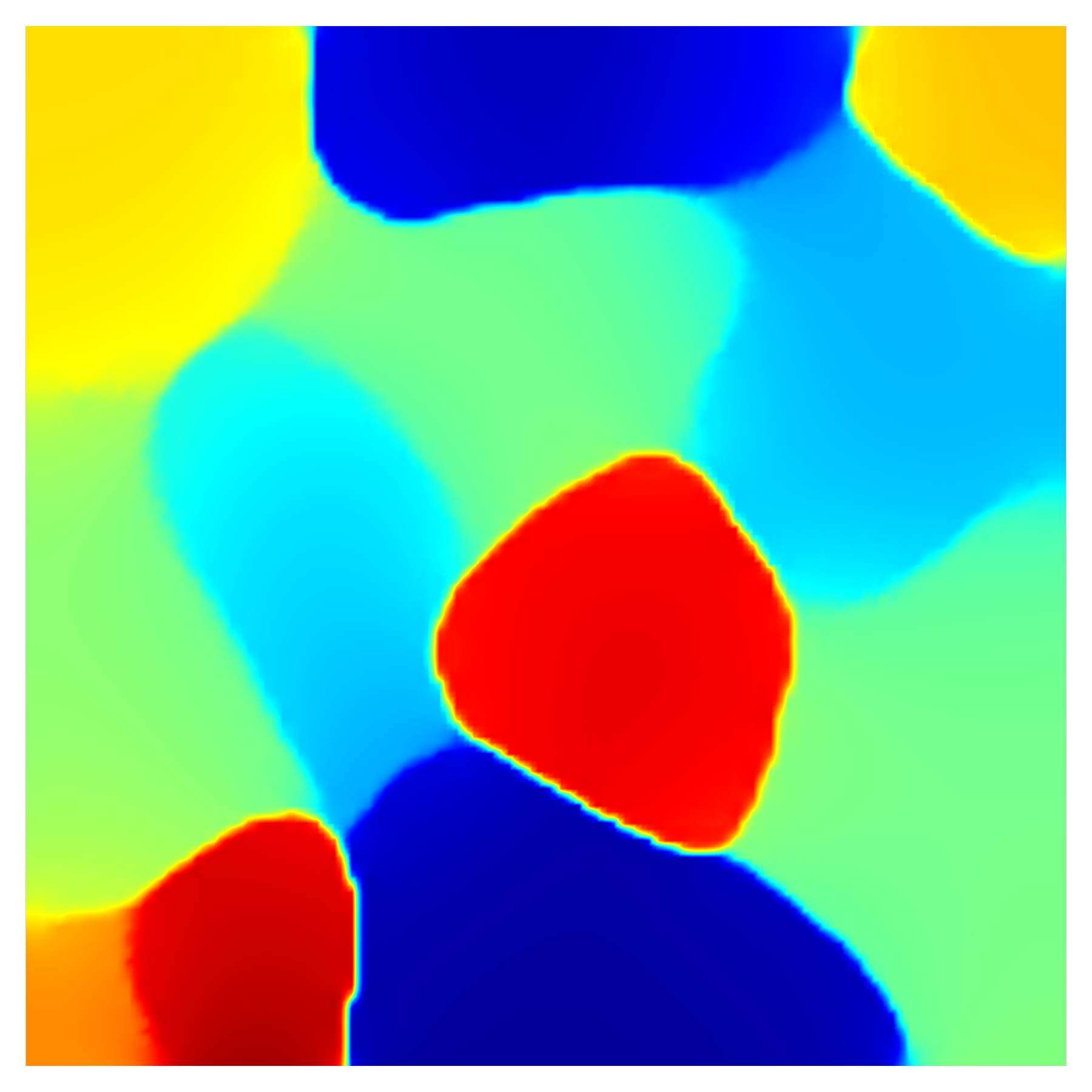}
        \label{fig:te_nl_2}}
    \end{minipage}
    &
    \begin{minipage}[c]{0.28\textwidth}
       \centering 
        \subfloat[$t=2\times10^{-5}$s]{\includegraphics[width=\textwidth]{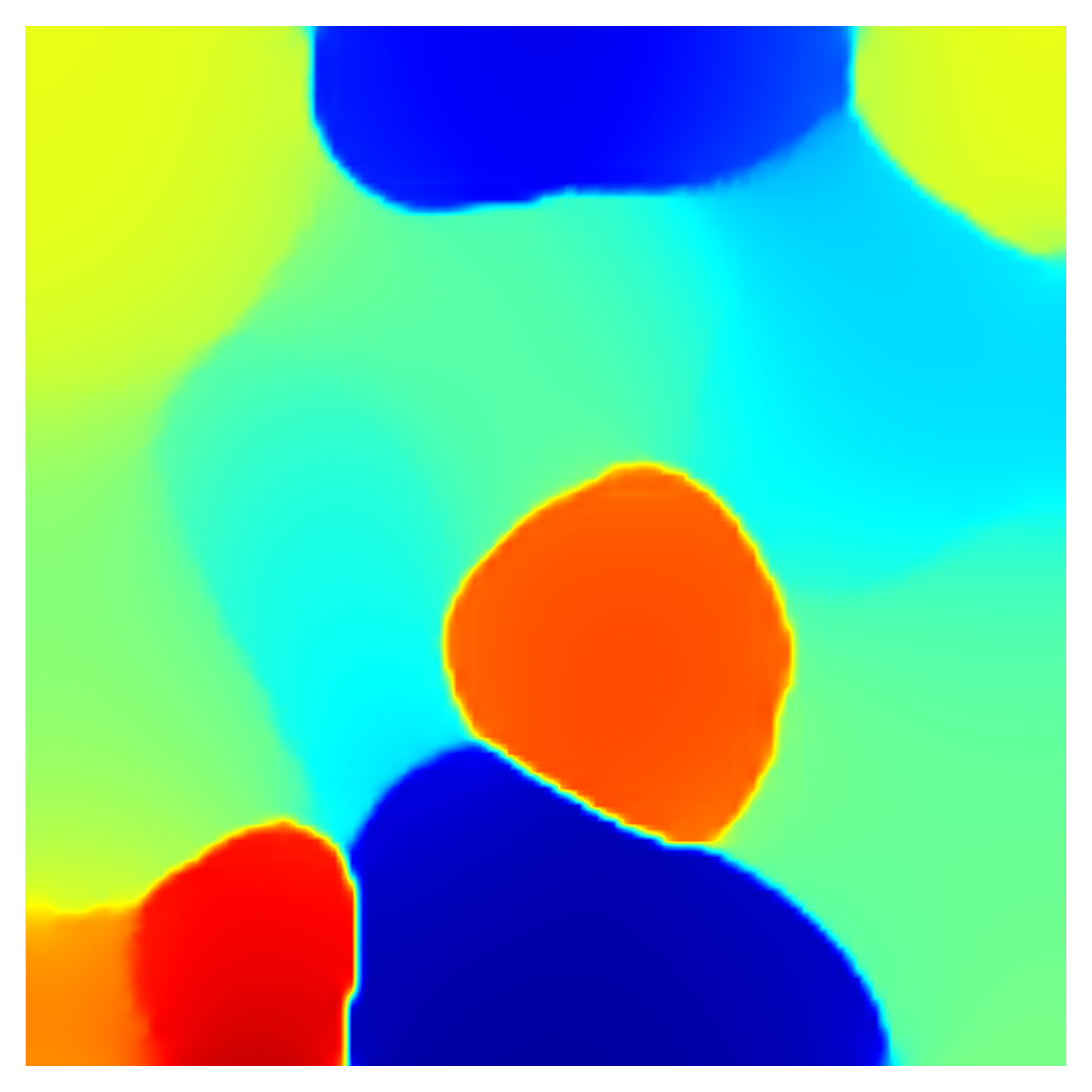}
        \label{fig:te_nl_3}}
    \end{minipage}\\

    \begin{minipage}[c]{0.28\textwidth}
       \centering 
        \subfloat[$t=3\times10^{-5}$s]{\includegraphics[width=\textwidth]{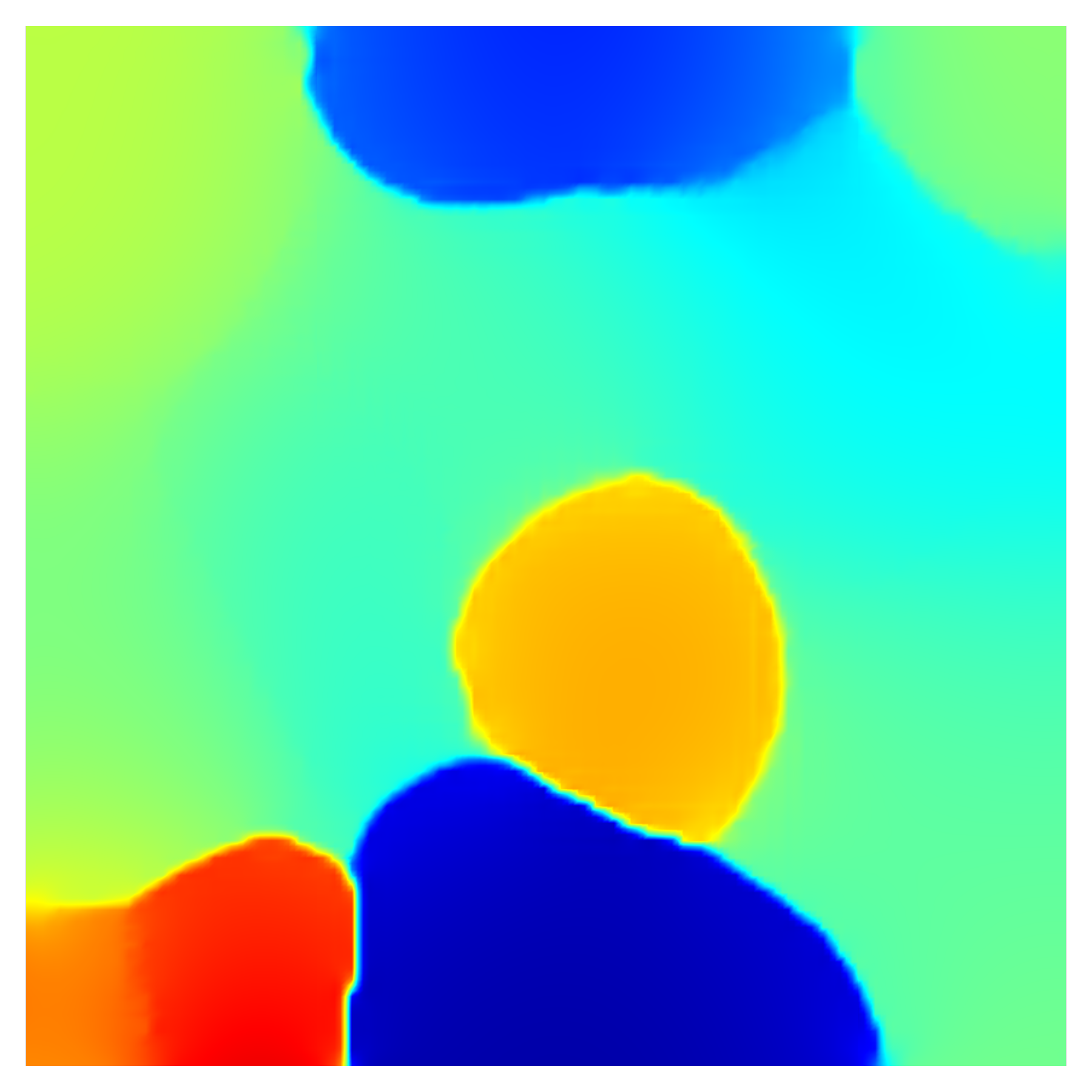}
        \label{fig:te_nl_4}}
    \end{minipage}
    &
    \begin{minipage}[c]{0.28\textwidth}
       \centering 
        \subfloat[$t=1\times10^{-4}$s]{\includegraphics[width=\textwidth]{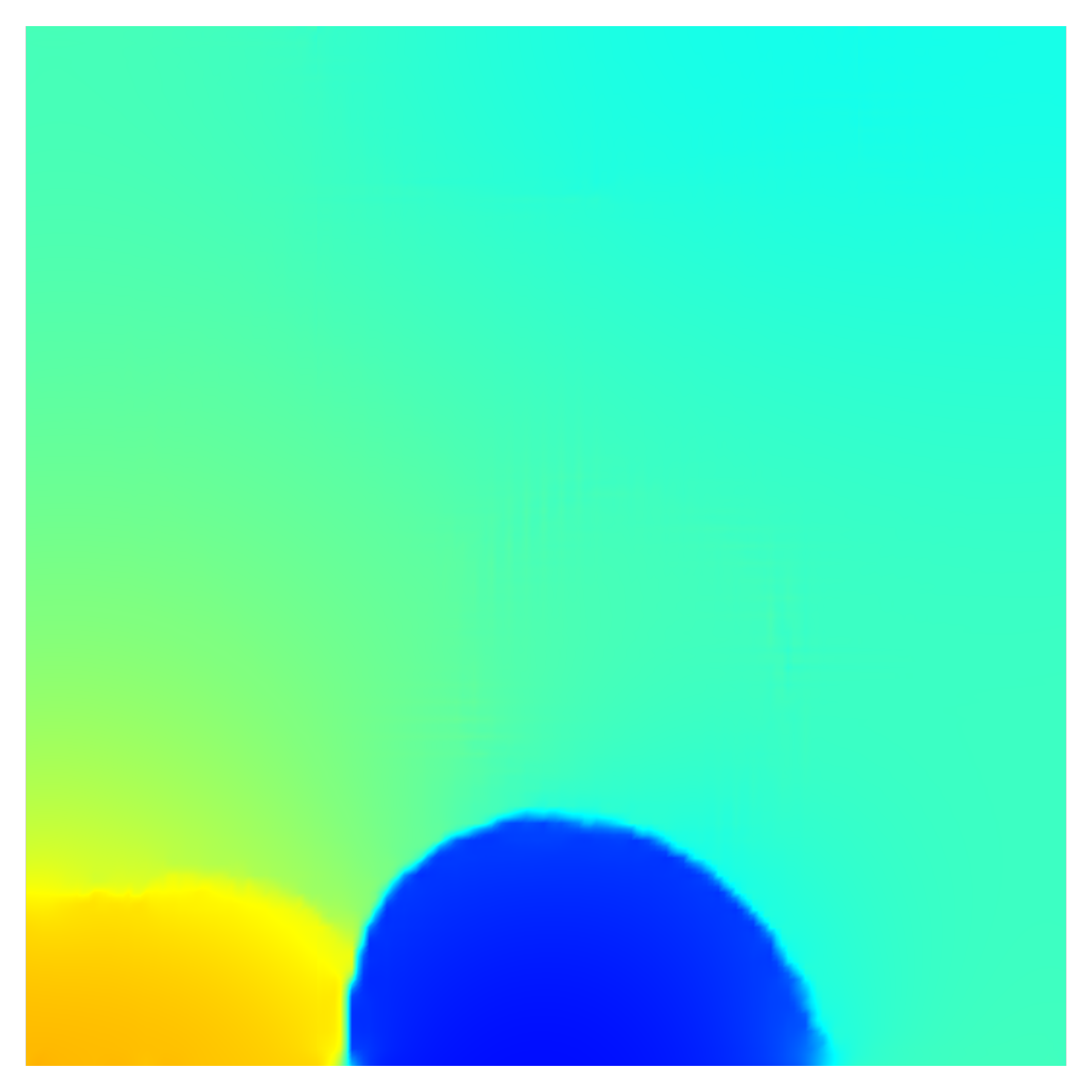}
        \label{fig:te_nl_5}}
    \end{minipage}
    &
    \begin{minipage}[c]{0.28\textwidth}
       \centering 
        \subfloat[$t=2\times10^{-4}$s]{\includegraphics[width=\textwidth]{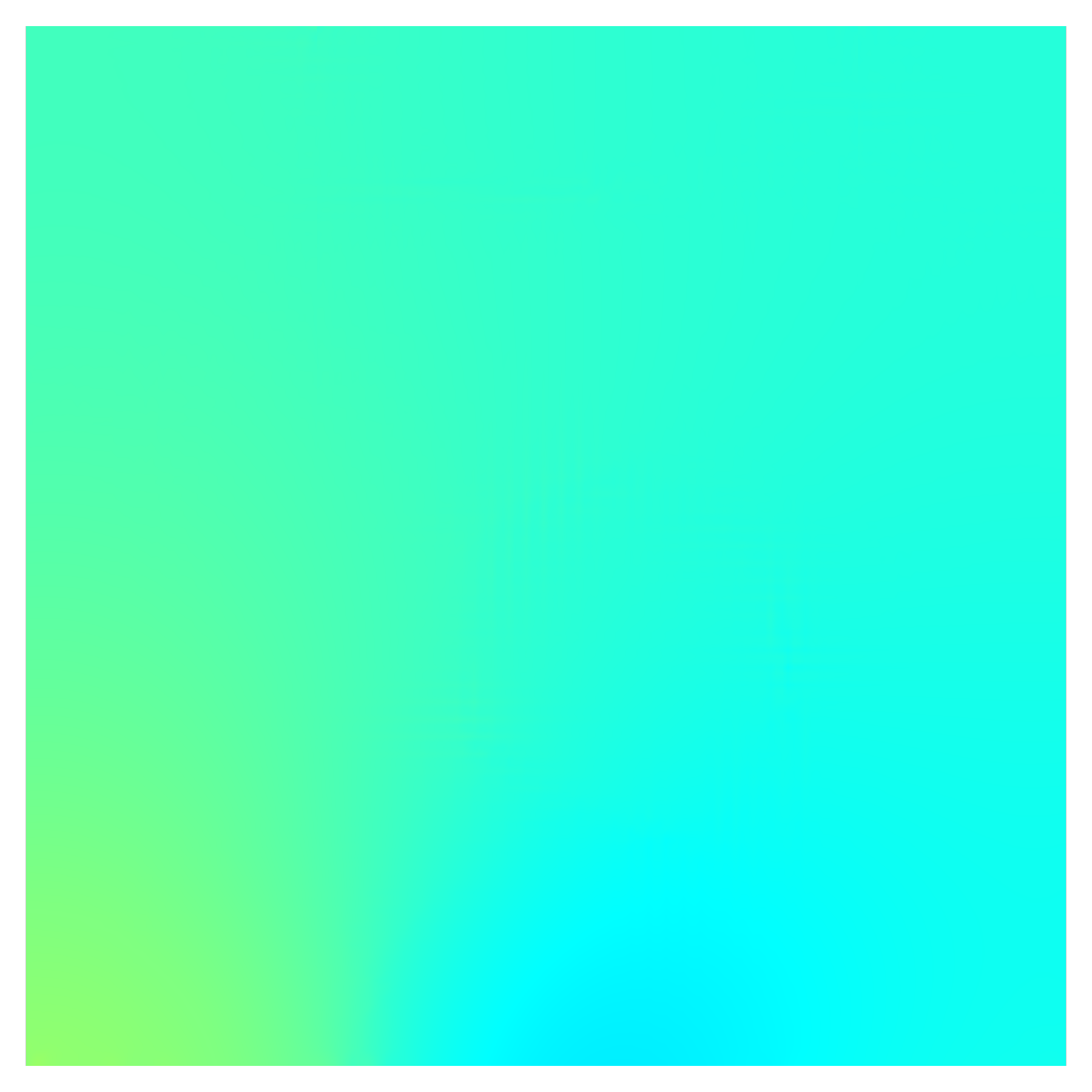}
        \label{fig:te_nl_6}}
    \end{minipage}
    & \\
    \end{tabular}
    \caption{Evolution of the polycrystal with no external loads: \psubref{fig:te_nl_1} initial condition with the grain labeling in \fref{fig:poly_ini} repeated, \psubref{fig:te_nl_2} $t=1\times10^{-5}$s, \psubref{fig:te_nl_3} $t=2\times10^{-5}$s, \psubref{fig:te_nl_4} $t=3\times10^{-5}$s, \psubref{fig:te_nl_5} $t=1\times10^{-4}$s, \psubref{fig:te_nl_2} $t=2\times10^{-4}$s. At the end of the simulation, the largest misorientation is less than 3.5$^{\circ}$, hence it is close to being a single crystal. Grains 1, 3 ,4 and 5, whose initial crystal orientation is higher than their neighbours, rotate to decrease their orientations. On the other hand, grain 2 rotates to increase its orientation.}
    \label{fig:PC_curv}
\end{figure}

\fref{fig:PC_curv} and \fref{fig:PC_stress} show snapshots (taken at the same time instances) of the time evolution of grains under no external load and uniaxial load cases, respectively. Clearly, in both cases, GB evolution is accompanied by grain rotation. However, we note that uniaxial loading has a noticeable impact on the rate of grain rotation. For instance, a comparison of \fref{fig:te_nl_4} and \fref{fig:te_wl_4} shows that while grain 1 rotates to decrease its misorientation with the neighbouring grains, the rate of lattice rotation is lower in the presence of external loads. This can be seen in \fref{fig:te_nl_4} wherein grain 1 ceases to exist by merging with its neighbor by $t=3\times10^{-5}$s, while it is still visible in \fref{fig:te_wl_4}. The above observation extends to grain 4 as well.

\begin{figure}[h!] 
    \centering
     \subfloat[]{
         \includegraphics[width=0.35\textwidth]{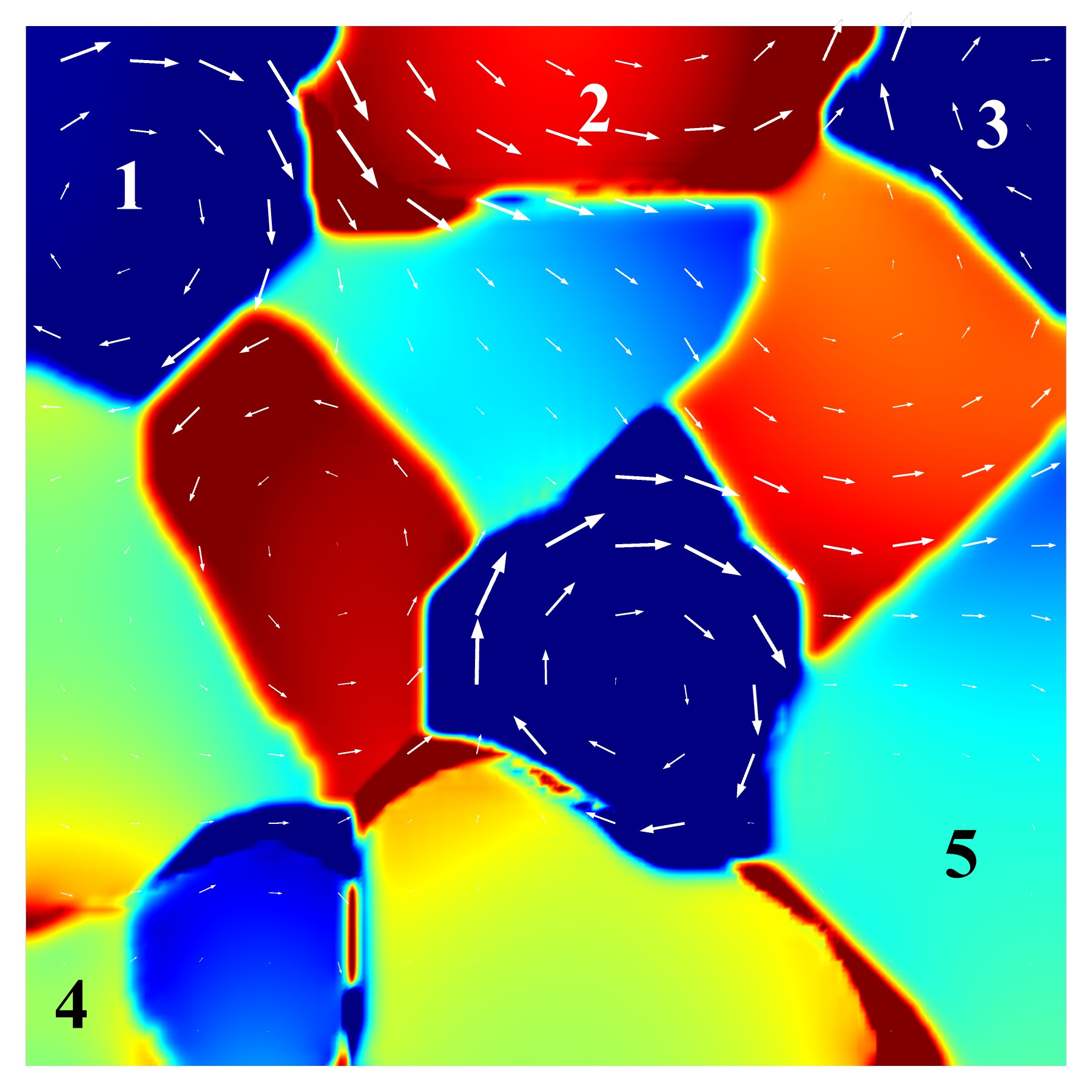}
         \label{fig:dte1}
     }
     \subfloat[]{
         \includegraphics[width=0.40833\textwidth]{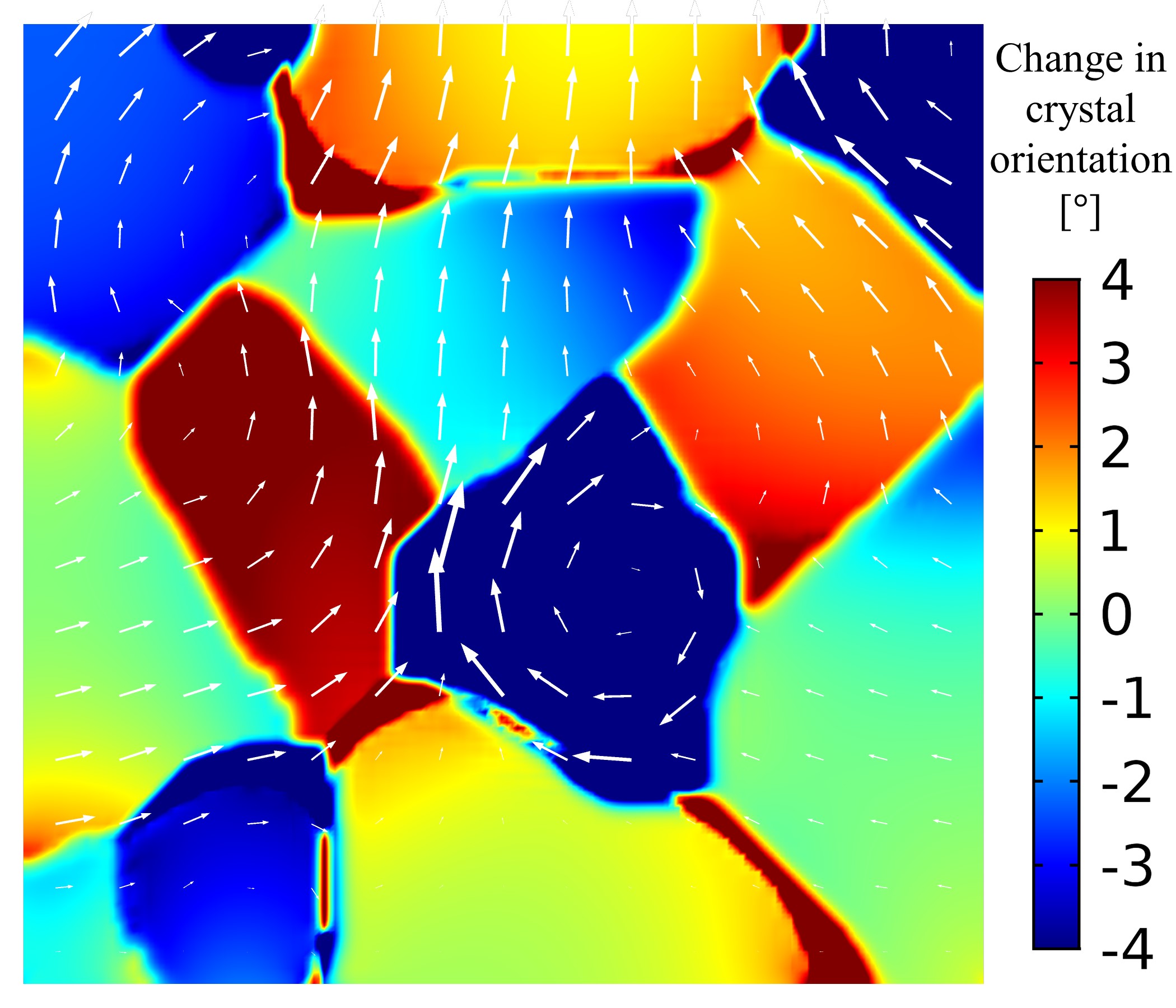}
         \label{fig:dte2}
     }
    \caption{Change in crystal orientation and displacement arrow surface plot at $t=3\times10^{-5}$s: \psubref{fig:dte1} With no external load,
    \psubref{fig:dte2} With applied displacement boundary conditions. Arrows are scaled with a uniform scale factor of 10. Comparing the two cases, it can be seen that the applied displacement changes the orientation evolution. Rotation of grains 1, 2, 5 is hindered, while rotation of grain 3 is enhanced. A negative rotation is also induced on grain 4 by the applied loads.}
    \label{fig:PC_curv_dte}
\end{figure}

The effect of external load on grain rotation is more evident in \fref{fig:PC_curv_dte}, which compares grain lattice rotations for the two load cases at time $t=3\times 10^{-5}$s. The arrows in \fref{fig:PC_curv_dte} depict displacement vectors. From \fref{fig:dte1}, we note that in the absence of external load, lattices of grains 1 and 3 rotate clockwise, while the lattice of grain 2 rotates counterclockwise. On the other hand, \fref{fig:dte2} shows that the presence of an external load on grain 2, superimposes a counterclockwise rotation in grain 1 and a clockwise rotation in grain 3. This results in a reduced lattice rotation for grain 1 while the lattice rotation is enhanced in grain 3. Moreover, lattice rotation of grain 2 is partially suppressed in the presence of external loads. 
\begin{figure}[h!]
    \centering
    \begin{tabular}{ c  c  c  c }
     & & & 
    \multirow{2}{*}{
    \begin{minipage}[c]{0.08\textwidth}
       \centering 
        \includegraphics[width=\textwidth]{ColorBar_PC.png}
    \end{minipage}
    }\\
    \begin{minipage}[c]{0.28\textwidth}
       \centering 
        \subfloat[$t=0$]{\includegraphics[width=\textwidth]{PC_ini_labeled.png}
        \label{fig:te_wl_1}}
    \end{minipage}
    &
    \begin{minipage}[c]{0.28\textwidth}
       \centering 
        \subfloat[$t=1\times10^{-5}$s]{\includegraphics[width=\textwidth]{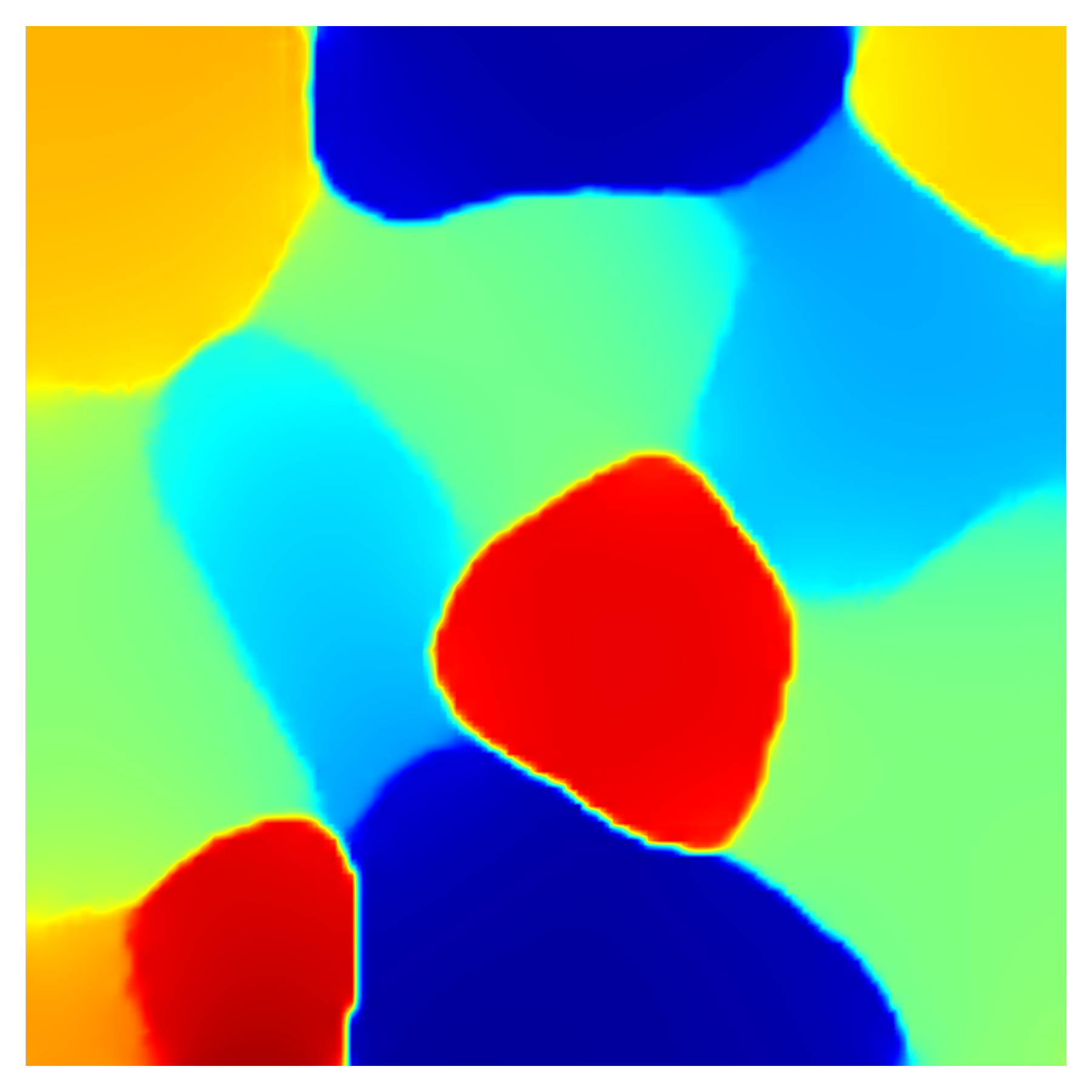}
        \label{fig:te_wl_2}}
    \end{minipage}
    &
    \begin{minipage}[c]{0.28\textwidth}
       \centering 
        \subfloat[$t=2\times10^{-5}$s]{\includegraphics[width=\textwidth]{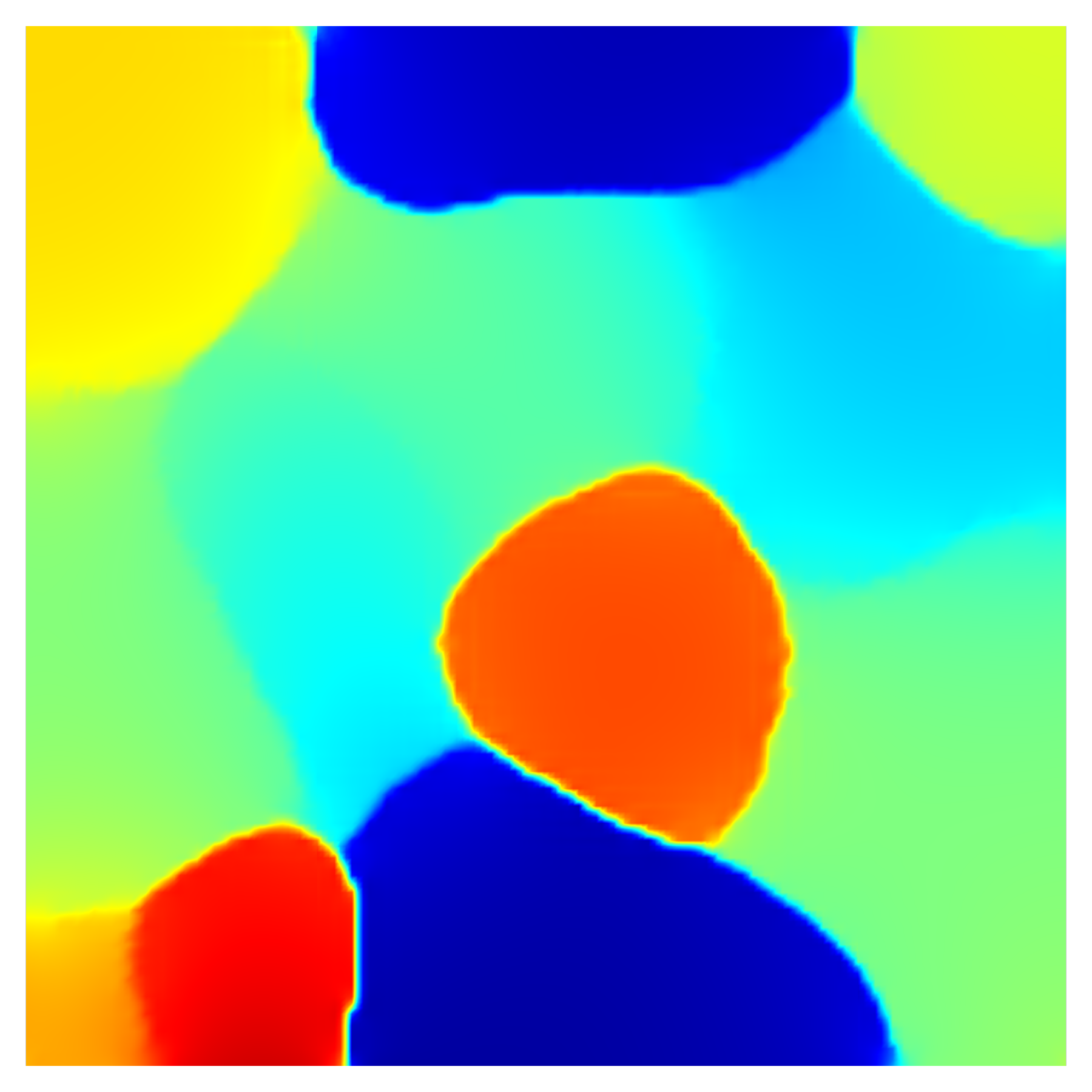}
        \label{fig:te_wl_3}}
    \end{minipage}\\

    \begin{minipage}[c]{0.28\textwidth}
       \centering 
        \subfloat[$t=3\times10^{-5}$s]{\includegraphics[width=\textwidth]{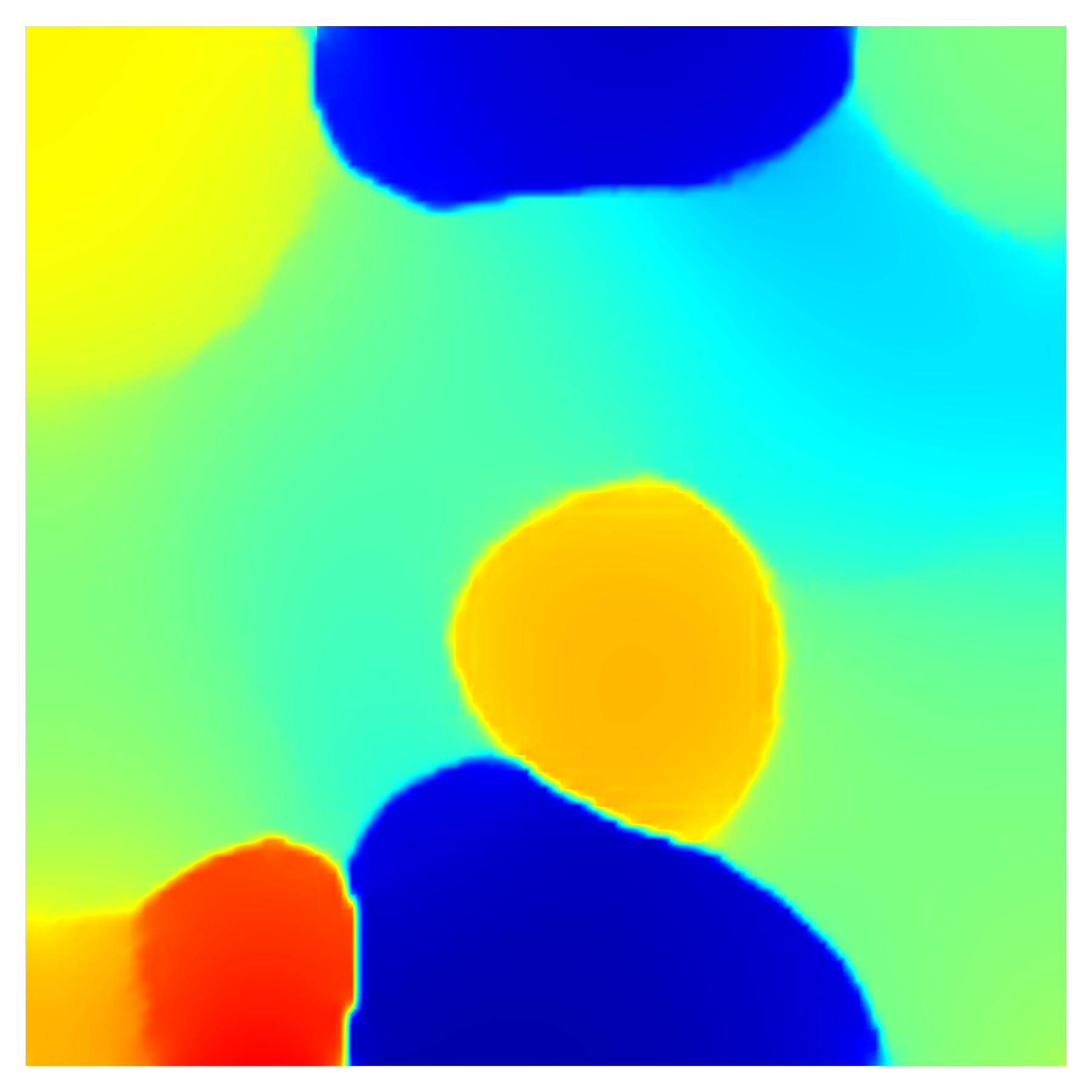}
        \label{fig:te_wl_4}}
    \end{minipage}
    &
    \begin{minipage}[c]{0.28\textwidth}
       \centering 
        \subfloat[$t=1\times10^{-4}$s]{\includegraphics[width=\textwidth]{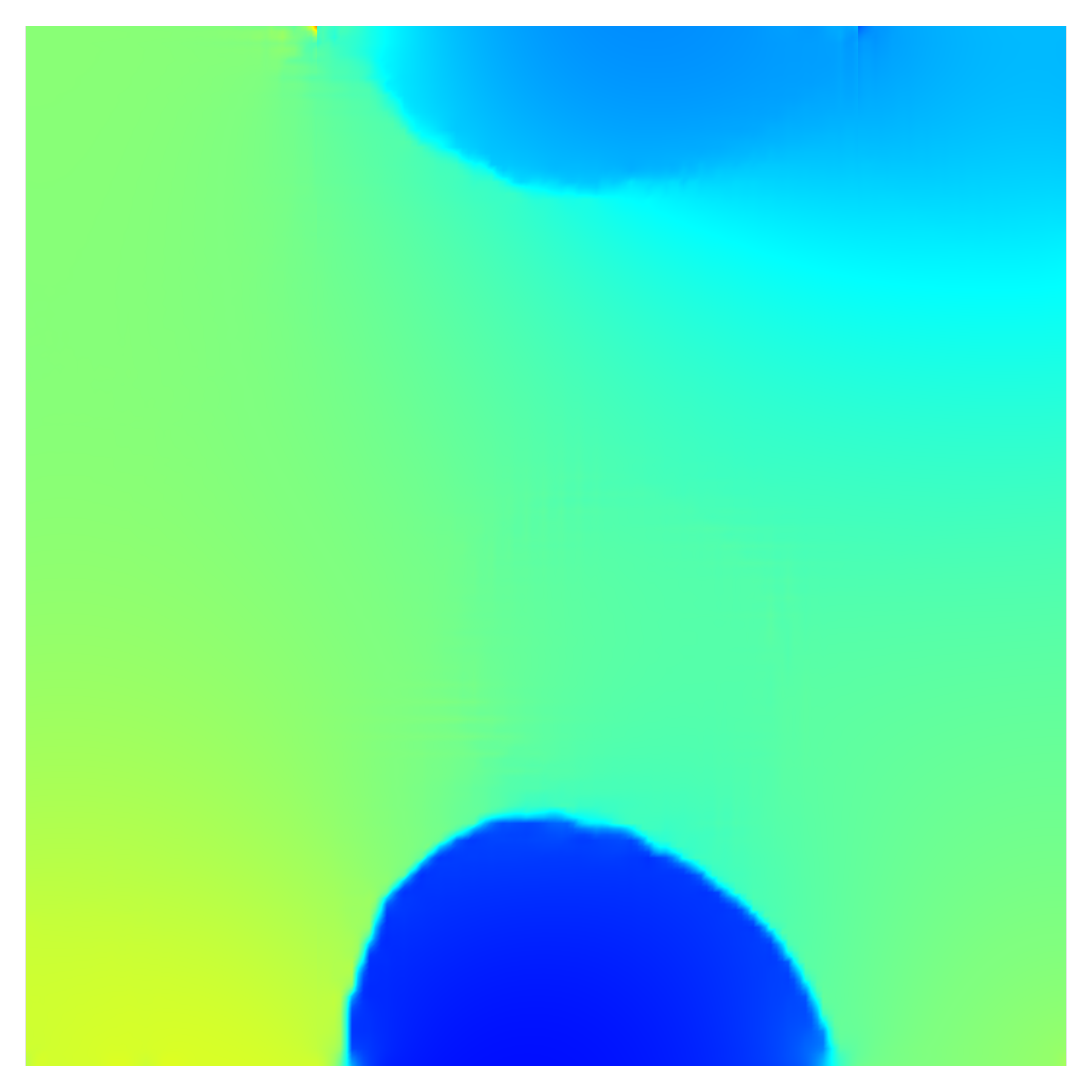}
        \label{fig:te_wl_5}}
    \end{minipage}
    &
    \begin{minipage}[c]{0.28\textwidth}
       \centering 
        \subfloat[$t=2\times10^{-4}$s]{\includegraphics[width=\textwidth]{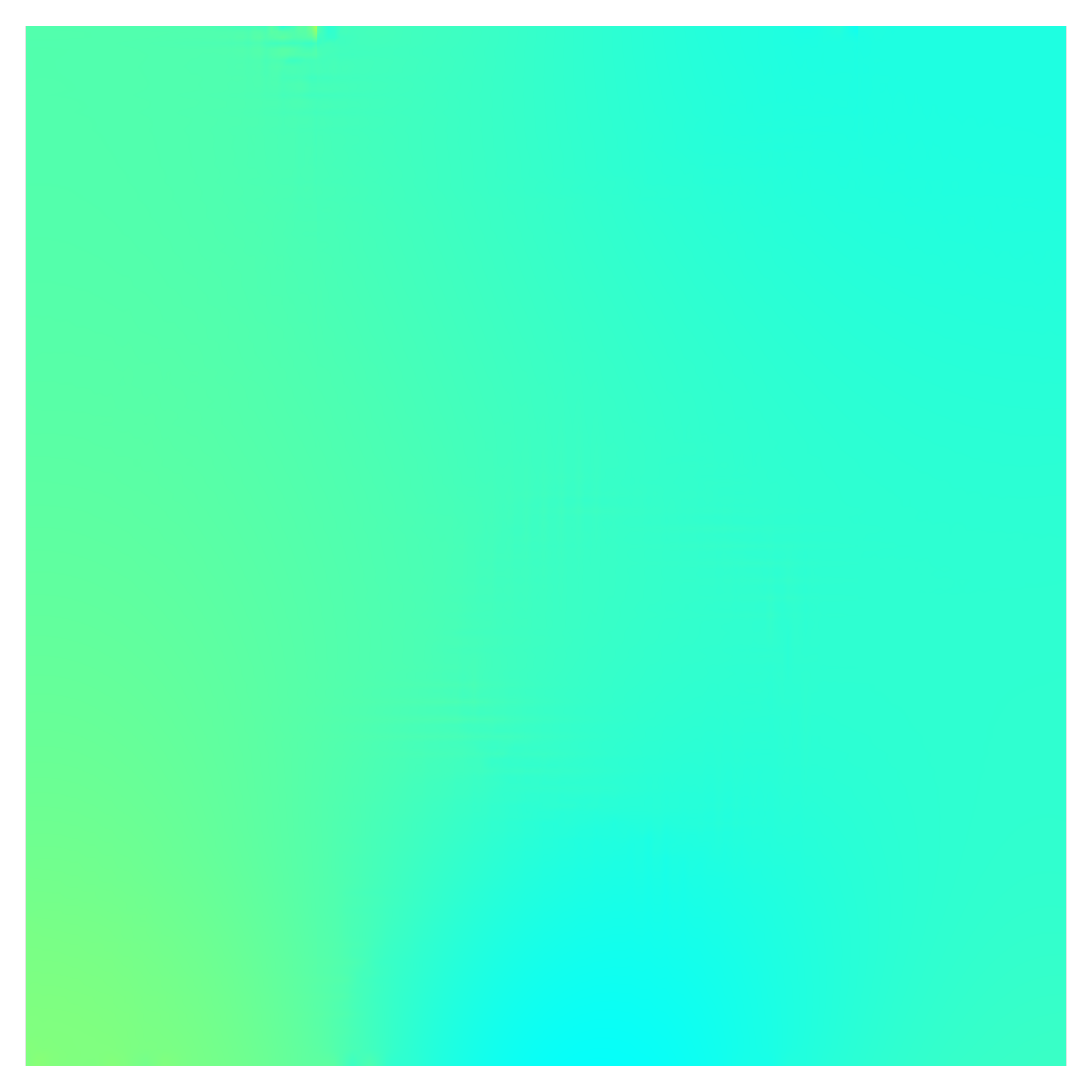}
        \label{fig:te_wl_6}}
    \end{minipage}
    & \\
    \end{tabular}
    \caption{Evolution of the polycrystal under applied loads: \psubref{fig:te_wl_1} initial condition with the grain labeling in \fref{fig:poly_ini} repeated, \psubref{fig:te_wl_2} $t=1\times10^{-5}$s, \psubref{fig:te_wl_3} $t=2\times10^{-5}$s, \psubref{fig:te_wl_4} $t=3\times10^{-5}$s, \psubref{fig:te_wl_5} $t=1\times10^{-4}$s, \psubref{fig:te_wl_2} $t=2\times10^{-4}$s. The external load slows down the negative rotation of grains 1 and 2, making them clearly distinguishable until up to $3\times10^{-5}$s. While the induced rotation on grain 4 causes it to merge with surrounding grains at $1\times10^{-4}$s.}
    \label{fig:PC_stress}
\end{figure}
In addition, additional rotations are introduced to grains 4 and 5, causing them to slightly rotate in the negative and positive direction, respectively.

In this simulation, equal mobilities are assigned to all available slip systems for simplicity. The work by \cite{thomas2017reconciling} highlights the importance of GB-specific migration behaviors in polycrystal evolution by the MD simulation of a simple toy polycrystal. As shown in \cite{admal2018unified}, the migration behavior of a GB (e.g., geometric coupling, sliding, or a mix of both) within this framework can be controlled numerically by the available slip systems. For example, it was shown that having two slip systems instead of three, would result in motion by curvature with no grain rotation. Within our framework, various GB mobilities can be explored by having a spatially varying field $c^\alpha$ (see \eref{eqn:b_alpha}), as guided by experimental measurements or individual MD simulations \cite{thomas2017reconciling}, thereby allowing different migration behaviors for GBs in the system. 

It is worth noting that the polycrystal simulation reveals a drawback of the diffuse-interface KWC-based models. During the evolution, as the misorientation of the GB decreases, the crystal orientation becomes more diffused. This can be seen in later stages of the two simulations (e.g., \fref{fig:te_nl_4} and \fref{fig:te_wl_4}). At these time instances, the crystal orientation field become diffused, which renders the identification of the location of GBs difficult. Note that similar diffused GBs can be seen in the polycrystal simulations by \cite{admal2019three}. This can be mitigated by increasing the value of $\gamma$ in \tref{tab:mat_prop}, which is used to approximate the singular-diffusive term $|\bm{G}|$ in \eref{eqn:phi}. However, doing so stiffens the resulting PDEs and worsens the conditioning of the global system, hence it is not pursued here. 

\section{Conclusion and discussion}
\label{sec:conclusion}
A computationally efficient framework for the simultaneous inclusion of grain boundary and bulk plasticity is proposed to simulate coupled polycrystal plasticity and grain boundary evolution. In this framework, which is based on the work of \citet{admal2018unified}, grain boundaries are represented by specialized arrangements of geometrically necessary dislocations that result in zero lattice strain, through the use of nonzero initial conditions for the lattice and plastic deformation gradients. This treatment encodes the grain orientation information into the plastic deformation gradient, and allows for the co-evolution of bulk and grain boundary plasticity using the evolution equation for the plastic deformation gradient in classical crystal plasticity theories. We introduced a synthetic driving potential within the framework, which allows us to drive the migration of a flat grain boundary without applying any mechanical loads, much like in molecular dynamics simulations. The primary difference between our framework and that of \citet{admal2018unified} is that we no longer solve for individual slip rates, which in unison, give rise to grain boundary plasticity. Instead, their combined effect is formulated as a second-order differential equation for plastic distortion. As a result, slip rates on each slip system are no longer nodal degrees of freedom that need to be solved for, and the number of degrees of freedom decreases, enhancing the computational efficiency.

Three simulations were conducted to demonstrate the new framework. In the first example, a scaling test is constructed to compare the computational speed of the two formulations, and it is revealed that the new formulation requires fewer degrees of freedom for the same mesh, and thus having shorter solution time. We also leverage this example to demonstrate the synthetic potential, to show that it indeed induces grain boundary migration on selected grains in the expected direction. For the second example, the evolution of a circular grain embedded in a bicrystal is studied. In the curvature-driven case, symmetric grain boundary migration is observed, and the area of the inner grain decreases linearly in time, both are in qualitative agreement with the MD results by \cite{trautt2014capillary}. When an external shear stress is applied, symmetry of the grain boundary system is broken, and negative rotation is observed on the circular grain, which is consistent with the MD results. The qualitative agreement with MD results offers confidence on the model's application to treat more complex problems. In the last numerical example, we demonstrate how an applied normal displacement changes the evolution of a polycrystal by introducing additional grain rotations on selected grains. The simulation predicts rotation directions that are consistent with a qualitative analysis. This example highlights the importance of external loads in the evolution of polycrystal texture, and motivates the use of this framework  to model phenomena in severe plastic deformation in large polycrystals. 

While the computational efficiency of our formulation will enable us to study
grain statistics of mechanically loaded polycrystals in the future, it is
important to note the limitations of our model. We start by noting some
limitations inherited from the model by \citet{admal2018unified}, which forms
the basis of the current work. The most prominent one is the dislocation
representation of GBs, which inherently limits the scope of our framework to
low-angle GBs due to the non-uniqueness of GNDs constituting the grain boundary,
which ultimately limits the energy of grain boundaries to the Ready--Shockley type. Therefore,
the current formulation is applicable to grain boundaries of misorientation
$<20^\circ$.
The above limitation can be partially remedied by using a lattice symmetric-invariant
KWC model, as in the work by \citet{admal:kim:2020}. Another limitation, also
rooting from the dislocation perspective of GBs, is that the sign of the
simulated coupling factor is fixed, as dictated by the choice of the GND
describing the GB. This is in contradiction with well-known observations
\citep{cahn2006coupling}, which find the coupling factor to not only be
multi-valued but can also change sign depending on the loading conditions.
In the current formulation, we assumed the inverse mobility for $\nabla \nu^\alpha$
to be a scalar multiple of the inverse mobility for $\nu^\alpha$.
This assumption was not present in the work of \citet{admal2018unified}, but was
necessary in the new formulation as we group all terms related to
$\nu^\alpha$ and $\nabla \nu^\alpha$ together. The implications of this
assumption is not clear, and is to be explored in future works, when a detailed
material model calibration is performed.

In order to address the limitations relating to high-angle GBs, we take note of the recent developments \citep{thomas2017reconciling,han2018grain} in interpreting GB motion at the atomic scale using disconnections. MD simulations \citep{han2018grain} demonstrate that disconnections are the primary carriers of GB plasticity in --- both low and high angle --- GBs with a well-defined coincident site lattice (CSL). This observation has motivated the recent development of disconnection-based continuum models \citep{wei2019continuum,chen2020grain}. While continuum disconnections are ideal candidates to model GB motion in high angle GBs, we note that the disconnections framework is only applicable to misorientations with a well-defined CSL lattice. Since a GB with an arbitrary misorientation can be interpreted as a "nearest" GB with a CSL lattice decorated with secondary dislocations\footnote{Secondary dislocations are disconnections which account for the deviation of the GB misorientation from a misorientation that results in a CSL lattice.}, our dislocations-based model will complement a disconnection-based model by modeling the secondary dislocations. This will be the focus of our future work.

Finally, we note that the low angle GBs play an important role in  phenomena such as abnormal grain growth, recrystallization and dynamic recovery. In addition, it has been well established that thermomechanical loading plays a key role in abnormal grain growth \cite{omori2013abnormal,moriyama2003effect,zielinski1995influence}, hence the application of our framework to explore these problems is also the focus of our future works.

\appendix
\section{Derivation of governing equations}
\label{sec:appendix}

To arrive at the constitutive equations, the energy balance is first examined. In the absence of any heat transfer, the energy balance reads:
\begin{equation}
    \dot{\epsilon} = (\bm{P}:\dot{\bm{F}} + \bm{p} \cdot \nabla\dot{\phi} + \pi \dot{\phi} ) + \sum_{\alpha=1}^{N_s}(\Pi^\alpha\nu^\alpha + \bm{\xi}^\alpha \cdot \nabla\nu^\alpha),
    \label{E:energy_balance}
\end{equation}
where $\epsilon$ denotes the energy density. For convenience, the stress power $\bm{P}:\dot{\bm{F}}$ is expressed in terms of the lattice Green-Lagrange strain and the slip rates:
\begin{equation}
    \bm{P}:\dot{\bm{F}} = \bm{S}:\dot{\bm{E}^{\rm{L}}} + \sum_{\alpha=1}^{N_s} \tau^\alpha \nu^\alpha,
    \label{E:stress_power}
\end{equation}
where the intermediate stress $\bm{S}=\bm{F}^{\rm L-1} \bm{P} \bm{F}^{\rm PT}$ and the resolved shear stress $\tau^\alpha=\bm{S}:(\bm{C}^{\rm L}\bm{S}^\alpha)$.

In this paper, since we limit to isothermal conditions, the Clausius--Duhem inequality is given by
\begin{equation}
    \dot\psi - \dot\epsilon \le 0.
\end{equation}
Using \eref{E:energy_balance}, the Clausius--Duhem inequality reads:
\begin{equation}
    \dot{\psi} - (\bm{S}:\dot{\bm{E}^{\rm{L}}} + \bm{p} \cdot \nabla\dot{\phi} + \pi \dot{\phi} ) - \sum_{\alpha=1}^{N_s}(\Pi^\alpha\nu^\alpha + \tau^\alpha\nu^\alpha + \bm{\xi}^\alpha \cdot \nabla\nu^\alpha) \le 0.
    \label{E:CD_2}
\end{equation}
Recalling the constitutive law for free energy from \sref{sec:equations}, the time derivative of $\psi$ can be expressed via the chain rule as
\begin{equation}
    \dot{\psi} = \frac{\partial\psi}{\partial\bm{E}^{\rm L}}:\dot{\bm{E}^{\rm{L}}} + \frac{\partial\psi}{\partial\phi}\dot{\phi} + \frac{\partial\psi}{\partial\nabla\phi}\cdot\nabla\dot{\phi} + \frac{\partial\psi}{\partial\bm{G}}:\dot{\bm{G}},\quad \text{where}
    \label{E:psi_dot}
\end{equation}
\begin{equation}
    \dot{\bm{G}} = \bm{L}^{\rm P}\bm{G} + \bm{G}\bm{L}^{\rm PT} + {\rm{det}}(\bm{F}^{\rm P}) \sum_{\alpha=1}^{N_s} \left[ (\bm{F}^{\rm P-T} \nabla\nu^\alpha) \times \bm{m}^\alpha \right] \otimes \bm{s}^\alpha.
    \label{E:G_dot}
\end{equation}
Inserting \eref{E:psi_dot} and \eref{E:G_dot} into \eref{E:CD_2}, we obtain 
\begin{equation}
  \begin{array}{l}
    (\frac{\partial\psi}{\partial\phi}-\pi)\dot{\phi} + (\frac{\partial\psi}{\partial\nabla\phi}-\bm{p})\cdot\nabla\dot{\phi}
    + \sum_{\alpha=1}^{N_s}( \frac{\partial\psi}{\partial\bm{G}}:(\bm{S}^\alpha\bm{G} + \bm{G}\bm{S}^{\alpha T}) - \tau^\alpha - \Pi^\alpha )\nu^\alpha \\
    + ( \frac{\partial\psi}{\partial\bm{E}^{\rm L}}-\bm{S}):\dot{\bm{E}^{\rm{L}}} + \sum_{\alpha=1}^{N_s}\left[ (\bm{F}^{{\rm P}T}\bm{m}^\alpha)\times(\bm{F}^{{\rm P}T}\frac{\partial\psi}{\partial\bm{G}}\bm{s}^\alpha) - \bm{\xi}^\alpha \right] \nabla\nu^\alpha \le 0.
  \end{array}
    \label{E:CD_detailed}
\end{equation}
The above inequality must be satisfied for all material points in the body. Using the Coleman--Noll procedure \citep{coleman1974thermodynamics}, we arrive at the relation
\begin{equation}
    \bm{S} = \frac{\partial\psi}{\partial\bm{E}^{\rm L}},
    \label{E:int_stress}
\end{equation}
as well as the relations given in \erefs{eqn:pk}{eqn:tau}.

\bibliographystyle{elsarticle-harv}
\setlength{\bibsep}{0.0pt}
{\footnotesize \bibliography{sample.bib} }

\end{document}